\documentclass[a4paper,11pt]{article}
\pdfoutput=1
\usepackage{jcappub}
\usepackage{inputenc}
\usepackage{mathtools}
\usepackage{dsfont}
\usepackage{xcolor}
\usepackage{float}
\graphicspath{{plots/}}
\usepackage[makeroom]{cancel}
\usepackage[normalem]{ulem}
\usepackage{slashed,soul,multirow, makecell}

\newcommand{\cmmnt}[1]{\ignorespaces}
\newcommand{\gs}{g_\star}
\newcommand{\gss}{g_{\star s}}

\newcommand{\Tfo}{T_\text{fo}}
\newcommand{\arh}{a_\text{rh}}
\newcommand{\afo}{a_\text{fo}}
\newcommand{\xfo}{x_\text{fo}}
\newcommand{\xrh}{x_\text{rh}}

\newcommand{\svk}{\langle\sigma_{r \to 2} v^{r-1} \rangle}

\title{SIMP dark matter during reheating}
\author[a]{ Debtosh Chowdhury,}
\author[a]{Sudipta Show}

\affiliation[a]{Department of Physics, Indian Institute of Technology Kanpur, Kanpur-208016, India}

\emailAdd{:debtoshc@iitk.ac.in}
\emailAdd{sudiptas@iitk.ac.in}

\abstract{Strongly interacting massive particle (SIMP) has become one of the promising dark matter (DM) candidates due to its capability of addressing the small-scale anomaly, where the final DM abundance is set via the freeze-out of  $3\rightarrow 2$ or $4\rightarrow 2$ annihilation process involving solely the dark sector particles. In this work, we explore the freeze-out of SIMP DM during the inflationary reheating epoch. During reheating, the radiation energy density evolves differently based on the shape of inflaton potential and spin of its decay products than the standard radiation-dominated picture; as a result, in this scenario, the freeze-out temperature varies distinctly with DM mass compared to the standard case. Large entropy injection due to inflaton decay demands a smaller cross-section to satisfy the observed relic than the standard radiation-dominated freeze-out case. The required cross-section, satisfying the relic density constraint and the maximum allowed thermally averaged cross-section by the unitarity of the $S$-matrix, set an upper limit on the DM mass. The upper bound on the mass of the dark matter for $3\rightarrow2$ ( $4\rightarrow2$ ) is $1$ GeV ($7$ MeV), assuming a radiation-dominated background. Interstingly, these limits get relaxed to $10^6$  ($10^4$) GeV for $3\rightarrow2$ ( $4\rightarrow2$ ) SIMP dark matter for quadratic inflaton potential. We find that a small amount of DM parameter space survives for reheating with quadratic inflaton potential after considering the lower bound of reheating temperature, put by the latest CMB observation depending on the inflationary models. In the case of the quartic inflaton potential, the allowed DM parameter space gets reduced compared to the quadratic case.}
\begin{document} 
\maketitle
\flushbottom
\section{Introduction}
Astrophysical and cosmological observations have conclusively demonstrated the prevalence of non-baryonic DM. In particular, cosmic microwave background (CMB) observations have established that dark matter contributes around 27$\%$ to the total energy budget of the universe and 80$\%$ to the total matter content, accounting for the relic density,$~\Omega h^2\simeq 0.12$~\cite{Planck:2018vyg, Drees:2018hzm}. Even though the identity of dark matter remains elusive, observations have unveiled some of its properties: it has to be electromagnetically neutral, colorless, collisionless (not strongly interacting), and non-relativistic, at least at the time of matter-radiation equality, needed for structure formation.

Numerous proposals have been put forth for dark matter candidates, and among them, the weakly interacting massive particles (WIMPs) are the most popular and well-explored scenario~\cite{Arcadi:2017kky}. However, despite decades of extensive efforts, conclusive evidence has yet to be found in the experiments. With this in mind, it is essential to pay attention to other avenues of dark matter such as the strongly interacting massive particle~\cite{Hochberg:2014dra, Ko:2014nha, Yamanaka:2014pva, Hochberg:2014kqa, Bernal:2015bla, Lee:2015gsa, Choi:2015bya, Hansen:2015yaa, Bernal:2015xba, Bernal:2015ova, BERNAL2015353, Heikinheimo:2016yds, Bernal:2017mqb, Choi:2017mkk, Heikinheimo:2017ofk, Choi:2017zww, Chu:2017msm, Bernal:2018ins, Bernal:2018hjm, Choi:2019zeb, Ho:2021ojb, Ho:2022erb, Ho:2022tbw, Parikh:2023qtk, Bernal:2024yhu}, Feebly Interacting Massive Paricle (FIMP) paradigm~\cite{Hall:2009bx, Bernal:2017kxu}, ELastically DEcoupling Relic (ELDER) scenario~\cite{Kuflik:2015isi, Kuflik:2017iqs, Bernal:2024yhu} etc. SIMPs are particularly noteworthy for their ability to potentially resolve longstanding issues in astrophysics related to small-scale structures such as the core-cusp problem~\cite{Flores:1994gz, Moore:1994yx, Oh_2011, Walker_2011} and the too-big-to-fail problem~\cite{Boylan_Kolchin_2011, Garrison_Kimmel_2014} etc., Owing to its large self-interaction at scales below a few Mpc. While dark matter strongly interacts with itself to achieve the relic density in the correct ballpark, it may weakly interact with the bath particles, ensuring it remains at least in kinetic equilibrium with the thermal bath before decoupling. Unlike WIMPs, where the dark matter relic density is set due to its interactions with the Standard Model (SM) particles, SIMP's number-changing process occurs entirely within the dark sector. In a minimal setup, the final abundance is decided by the freeze-out of the $3\rightarrow2$ process., where the DM number-changing reaction involves only DM particles. However, the dominant process depleting the number of particles can also be $4\rightarrow2$~\cite{Bernal:2015xba}, where symmetry prohibits the $3\rightarrow2$ interaction.

Generally, the radiation-dominated background in the early universe is considered while studying the production of dark matter. Till now, we only know that the universe was radiation-dominated at the time of Big Bang nucleosynthesis (BBN). However, there is no direct evidence of the energy budget before BBN. Therefore, it is essential to investigate the impact of alternative cosmologies on the evolution of DM. The study of dark matter decoupling during the non-standard expansion epoch has received growing interest~\footnote{Baryogenesis has been extensively explored in the low reheating temperature or early matter era; see refs~\cite{Konar:2020vuu, Davidson:2000dw, Allahverdi:2010im, Allahverdi:2017edd}. Additionally, studies on the generation of primordial gravitational waves during early matter-dominated epochs have gained particular focus~\cite{Assadullahi:2009nf, Alabidi:2013lya, DEramo:2019tit, Bernal:2019lpc, Figueroa:2019paj}.}; see refs~\cite{Co:2015pka, Davoudiasl:2015vba, Tenkanen:2016jic, DEramo:2017gpl, Hamdan:2017psw, Visinelli:2017qga, Drees:2017iod, DEramo:2017ecx, Maity:2018dgy, Bernal:2018ins, Hambye:2018qjv, Bernal:2018kcw, Arbey:2018uho, Drees:2018dsj, Poulin:2019omz, Arias:2019uol, Bernal:2019mhf, Mahanta:2019sfo, Chanda:2019xyl, Konar:2020vuu, Bhatia:2020itt, Barman:2021ifu, Ghosh:2021wrk, Banerjee:2022fiw, Das:2023owa, Bernal:2023ura, Bernal:2024yhu, Bernal:2024yhu, Banerjee:2024caa, Silva-Malpartida:2024emu}. 
Although WIMP dark matter during reheating has been widely explored in numerous works in the literature~\cite{Drees:2017iod, Gelmini:2006pw, Giudice:2000ex, Fornengo:2002db, Drees:2006vh, Roszkowski:2014lga, Bernal:2022wck}, there is no dedicated and detailed work on SIMP during reheating.

We investigate the phenomenology of SIMP dark matter, where the freeze-out of dark matter occurs during the reheating phase after inflation. We consider that the inflaton ($\phi$) oscillates around the minimum of a generic potential ($V(\phi)\propto \phi^k$). Potential with such behavior can be found in various inflationary models like $\alpha-$ attractor $T-$ and $E-$ models~\cite{Kallosh:2013hoa, Kallosh:2013maa, Starobinsky:1980te}, which have received support from recent observations~\cite{Planck:2018jri, BICEP:2021xfz, Kallosh:2021mnu}. Notably, the shape of the potential determines the equation of state of the inflation,  $\omega=(k-2)/(k+2)$~\cite{Turner:1983he}. Furthermore, both the mass and the decay width develop time (or field $\phi$) dependency~\cite{Garcia:2020eof, Garcia:2020wiy} for $k>2$. This differs from the typical quadratic potential (i.e., $k=2$ case), where the inflation mass and decay width are time-independent. Consequently, in such reheating scenarios, the energy density of radiation and inflaton exhibit non-trivial evolution, which also depends on the spin of the inflaton decay products provided $k>2$. Freeze-out during reheating allows for the exploration of much smaller thermally averaged cross-sections to counteract the effect of entropy injection caused by the inflaton decay, thus relaxing the unitarity constraint on thermal DM.

Dark matter has a wide range of mass spectrum.   The stability of stellar clusters in galaxies and the requirement de-Broglie wavelength to confine the dark matter (DM) within galaxies set the upper limit ($10^3M_\odot$)~\cite{Moore:1993sv, Carr1999DynamicalCO} and lower bound ($10^{-22}~\text{eV}$ for bosonic DM)~\cite{Hu:2000ke, Hui:2016ltb} respectively. The specification of specific properties further restricts such model-independent limits. For instance, for dark matter maintaining thermal equilibrium with the thermal soup in the early universe, the requirement of $S$-matrix unitarity sets an upper bound on mass approximately $\mathcal{O}(100)$ TeV~\cite{PhysRevLett.64.615, Hui:2001wy}, provided DM pair annihilates to SM particles. Likewise, for thermal dark matter, such model-independent unitarity upper limits are  $\mathcal{O}(1)$ GeV and $\mathcal{O}(1)$ MeV while the dominant number-changing processes are $3\rightarrow2$ and $4\rightarrow2$ respectively~\cite{Bhatia:2020itt, Bernal:2023ura}. The maximum allowed inelastic cross-section implied by $S$-matrix unitarity establishes DM's minimum frozen-out number density and, consequently, the maximum mass saturates the present density. All the bounds mentioned above are usually calculated for radiation-dominated backgrounds. However, alternative background cosmology modifies the required cross-section compared to the standard radiation-dominated case, consequently altering the unitarity upper bounds on DM mass. The study of unitarity mass bound on dark matter in non-standard cosmologies has been discussed in refs~\cite{Bhatia:2020itt, Bernal:2023ura}.

In our work, we investigated the effect of reheating on the evolution of SIMP dark matter for both quadratic and quartic inflationary potentials, considering inflaton decays to pairs of fermions and bosons. Our study revealed that the freeze-out temperature exhibits distinct dependencies on DM mass in the radiation-dominated background, fermionic, and bosonic reheating pictures. However, the freeze-out during reheating allows us to explore a wide range of smaller cross-sections than the standard scenario due to entropy injection; a lower limit exists on the cross-section for a given reheating temperature to ensure the thermal equilibrium of DM with the thermal bath. Thus, a lower cut-off on cross-section rules out some areas of parameter space. Moreover, the reheating temperature below BBN temperature conflicts with observations, which also restricts certain regions of the parameter space. The constraints of thermalization and the unitarity limit collectively place an upper bound on the dark matter mass, which gets significantly relaxed in the non-standard scenario. We also highlighted that the allowed DM parameter space is either partially or entirely excluded by the inflation model dependent lower bound on the reheating temperature, as dictated by the CMB experiments, for quadratic inflaton potential. In contrast, in the case of quartic potential, the DM picture remains free of the inflationary constraint since the reheating temperature is unbounded here.   

The outline of this work is as follows. We briefly review the inflationary reheating picture in section~\ref{sec2}. We demonstrate the Boltzmann equation for $r(\ge 3)\rightarrow 2$ annihilation of dark matter and the maximum allowed cross-section for this process by $S$-matrix unitarity in section~\ref{sec3}. Sections~\ref{sec4} and ~\ref{sec5} display the analytical expressions of the freeze-out and the thermally averaged cross-section for radiation-dominated background and fermionic and reheating pictures, respectively. We discuss our results and show the unitarity and inflationary constraints on DM parameter space in section~\ref{sec6}. Finally, we summarize our findings in section~\ref{sec7}.  

\section{Inflationary Reheating}\label{sec2}
We consider the scenario where the inflaton field($\phi$) is oscillating around the minimum of the following monomial potential
\begin{align}\label{Inf_Reh1}
V(\phi)=\lambda\frac{\phi^k}{\Lambda^{k-4}}
\end{align}
with $\lambda$ refers to dimensionless coupling and $\Lambda$ is the cut-off scale. Without loss of generality, we will consider this cut-off energy scale to be reduced Planck mass $M_P$($=2.4\times 10^{18}$ GeV). The equation of state parameter of $\phi$ is expressed as, $\omega\equiv p_\phi/\rho_\phi=(k-2)/(k+2)$~\cite{Turner:1983he} where $p_\phi(\equiv\frac{1}{2}\dot\phi^2-V(\phi))$ and $\rho_\phi(\equiv\frac{1}{2}\dot\phi^2+V(\phi))$ are the pressure and energy density of the inflaton. 
Being the dominant energy component during reheating, the inflaton energy density drives expansion where the Hubble rate can be cast as
\begin{equation}\label{Inf_Reh2}
H(a) \simeq
\begin{dcases}H_{\text{rh}}
\left (\frac{\arh}{a}\right)^{\frac{3k}{k+2}} & \text{ for } a < a_{\text{rh}},\\H_{\text{rh}}
\left(\frac{a_{\text{rh}}}{a}\right)^2 & \text{ for } a \geq a_{\text{rh}}.
\end{dcases}
\end{equation}
where $a$ and $a_{\text{rh}}$ are the scale factor at a time, t, and the same at the end of reheating, respectively. $H_{\text{rh}}$ represents Hubble rate at  $T_{\text{rh}}$ which is the bath temperature when the energy density of radiation and inflaton equals (i.e., $\rho_R (T_{\text{rh}})=\rho_\phi(T_{\text{rh}})$). Note that the reheating temperature ($T_{\text{rh}}$) must satisfy $T_{\text{rh}}\ge T_{\text{BBN}}(\simeq 4~\text{MeV})$~\cite{Sarkar:1995dd, Kawasaki:2000en, Hannestad:2004px, DEBERNARDIS2008192, deSalas:2015glj}, since we do not want to spoil the success of BBN. Here, one can calculate the effective mass of the inflaton field by taking the double derivative of the inflaton potential as\footnote{Envelope approximation is used for $\rho_\phi$, defined by $V(\phi_0(t))=\rho_\phi(t)$~\cite{Garcia:2020eof, Garcia:2020wiy}. The oscillating inflaton can be approximated as $\phi(t)\simeq\phi_0(t)\mathcal{P}(t)$ where the periodic function $\mathcal{P}(t)$ describes the an(harmonicity) on short time scales and the envelope $\phi_0(t)$ varies in longer time scale dictating the redshift and decay.} 
\begin{align}\label{Inf_Reh3}
m_\phi^2=\frac{d^2V}{d\phi^2}=k(k-1)\lambda\frac{\phi^{k-2}}{\Lambda^{k-4}}\simeq k(k-1)\lambda \Lambda^{\frac{2(4-k)}{k}}\rho_\phi(a)^{\frac{k-2}{k}}
\end{align}
with $\rho_\phi(a)$ is the energy density of the inflaton. The inflaton mass is field- or scale-factor dependent for $k>2$.

In this work, we study the DM evolution during reheating with low reheating temperature\footnote{Note that our analysis assumed instantaneous thermalization, the implications of non-instantaneous thermalization could be found in refs~\cite{Mukaida:2015ria, Garcia:2018wtq, Chowdhury:2023jft}.} where the inflaton field decays to two fermions or bosons via a trilinear coupling. In this case, the decay width ($\Gamma_\phi$) is proportional to the mass of the inflaton field for fermionic final states ($\Gamma_\phi^f\propto m_\phi$) and inversely to the same for bosonic final states ($\Gamma_\phi^b\propto 1/ m_\phi$). For $k>2$, the decay width shows different dependencies on the scale factor depending on the nature of the final state particles. Moreover, the evolution of the radiation energy density, $\rho_R(a)$, has a distinct behavior since it depends on the inflaton decay width. Once the radiation energy density is known, the temperature of the bath ($T(a)\propto \rho_R(a)^{1/4}$) can be calculated, and it also features different evolution for these two reheating scenarios. Next, we will discuss fermionic and bosonic reheating pictures one by one.

\subsection{Fermionic Reheating}
The evolution of the SM temperature when the decay products of the inflaton are fermions is given by
\begin{equation}\label{Inf_Reh4}
T(a) \simeq T_{\text{rh}} \times
\begin{dcases}
\left(\frac{a_{\text{rh}}}{a}\right)^{\frac{3}{2}\frac{k-1}{k+2}} & \text{ for } T > T_{\text{rh}}~~~~(k < 7),\\
\left(\frac{g_{\star s}(T_{\text{rh}})}{g_{\star s}(T)}\right)^{1/3} \frac{a_{\text{rh}}}{a} & \text{ for } T \leq T_{\text{rh}}.
\end{dcases}
\end{equation}
Here, $g_{*s}$ refers to the effective number of entropic degrees of freedom. We always consider $k<7$ throughout our analysis\footnote{For $k>7$, the temperature evolves as $T\propto a^{-1}$, so the Hubble rate also varies differently compared to the case with $k<7$\cite{Bernal:2022wck}.}. Now, the Hubble rate can be re-expressed in terms of bath temperature by using the Eq.~(\ref{Inf_Reh2}) as
\begin{equation}\label{Inf_Reh5}
H(T) \simeq
\begin{dcases}
H_R(T_{\text{rh}}) \left(\frac{T}{T_{\text{rh}}}\right)^{\frac{2k}{k-1}} & \text{ for } T > T_{\text{rh}},\\
H_R(T) & \text{ for } T \leq T_{\text{rh}}.
\end{dcases}
\end{equation}
where $H_R(T)~(\equiv\sqrt{\frac{\rho_R}{3M_P^2}}=\frac{\pi}{3}\sqrt{\frac{g_\star(T)}{10}}\frac{T^2}{M_P})$ denotes the Hubble rate during the radiation-dominated phase of the universe.

\subsection{Bosonic Reheating}
On the other hand, if the inflaton decays into a pair of bosons, the SM temperature evolves as
\begin{equation}\label{Inf_Reh6}
T(a) \simeq T_\text{rh} \times
\begin{dcases}
\left(\frac{a_\text{rh}}{a}\right)^{\frac{3}{2}\frac{1}{k+2}} & \text{ for } T > T_\text{rh},\\
\left(\frac{g_{\star s}(T_\text{rh})}{g_{\star s}(T)}\right)^{1/3} \frac{a_\text{rh}}{a} & \text{ for } T \leq T_\text{rh}.
\end{dcases}
\end{equation}
Following Eq.~(\ref{Inf_Reh2}), the Hubble expansion rate is
\begin{equation}\label{Inf_Reh7}
H(T) \simeq
\begin{dcases}
H_R(T_\text{rh}) \left(\frac{T}{T_\text{rh}}\right)^{2k} & \text{ for } T > T_\text{rh},\\
H_R(T) & \text{ for } T \leq T_\text{rh}.
\end{dcases}
\end{equation}
Note that, in quadratic potential ($k=2$ case) where the equation of state parameter $\omega=0$, the standard dependence with the scale factor $T(a)\propto a^{-3/8}$ is reproduced for both the cases.

\section{SIMP Paradigm: DM Evolution and Unitarity Bound on Cross-Section}\label{sec3}
The dominant number-changing process for SIMP dark matter is $3\rightarrow 2$ or $4\rightarrow 2$ when symmetry restricts $3\rightarrow 2$ interaction. In the case of SIMP, the interaction involves only the dark matter particles, unlike the WIMP scenario where the dominant $2\rightarrow 2$ process involves both DM and SM particles. Now, the evolution of the DM number density, $n(t)$ for a general $r\rightarrow 2$ number changing process with $r>2$, where all $r+2$ interacting particles are dark matters,  is governed by the following Boltzmann equation~\cite{Bhatia:2020itt}
\begin{align}\label{Evo_1}
\frac{dn}{dt} + 3\, H\, n = -(\Delta n)\svk \left(n^r - n^2\, n_\text{eq}^{r-2}\right),
\end{align}
with the Hubble rate
\begin{align}\label{Evo_2}
H(T)=\sqrt{\frac{\rho_\phi+\rho_R}{3M_P^2}},
\end{align}
where $\langle\sigma_{r\rightarrow 2} v^{r-1}\rangle$ is the thermally averaged annihilation cross-section of DM for $r\rightarrow 2$ process, $v$ refers to relative velocity of each particle pair, $\Delta n$ denotes the net change of DM numbers and $n_{\text{eq}}$ represents the DM equilibrium number density. We have not considered the contribution of DM matter energy density to the total energy density since it does not change the overall picture presented in this work.\\
The number density of the DM is tracked by solving Eq.~(\ref{Evo_1}) for a fixed DM mass,  $m$, and background cosmology, where the non-relativistic freeze-out fixes the final DM density. Here, we assume that the freeze-out always occurs when the dark matter is in thermal equilibrium with the visible particles. Now, one can obtain the value of thermally averaged cross-section for any DM mass satisfying relic density constraint by numerically solving Eq.~(\ref{Evo_1}). In addition, one can also deduce the analytical expression for the required thermally averaged cross-section featuring correct order DM relic by using Eq.~(\ref{Evo_1}). Then, one needs to know the freeze-out temperature, $T_\text{fo}$, to calculate the DM annihilation cross-section. The temperature $T_{\text{fo}}$ delineates the point after which DM deviates from the chemical equilibrium. The freeze-out temperature can be evaluated by using the equality between the Hubble rate and the dark matter annihilation rate
\begin{align}\label{Evo_5}
H(\Tfo) = n_\text{eq}^{r-1}(\Tfo)\, \svk,
\end{align}
In this scenario, the freeze-out occurs when the DM is non-relativistic, so one can solve the above equation using the equilibrium DM number density in the non-relativistic limit
\begin{align}\label{Evo_3}
n_{\text{eq}}(T)=g {\bigg(\frac{mT}{2\pi}\bigg)}^{3/2}e^{-\frac{m}{T}},
\end{align}
where $g$ is the internal degrees of freedom. However, one can obtain the required cross-section for DM solving Eq.~(\ref{Evo_1}) as mentioned earlier, but one needs to check the validity of it. Since unitarity of $S$-matrix restricts cross-section from above. 
The upper bound on the thermally averaged annihilation cross-section, placed by $S$-matrix unitarity, for $r\rightarrow 2$ process with $r\ge 2$ is given by~\cite{Bhatia:2020itt}
\begin{equation}\label{Evo_4}
\svk_\text{max} = \sum_l (2 l + 1)\, \frac{2^\frac{3r-10}{2}\, (\pi x)^\frac{3r-7}{2}}{g^{r-2}\, m^{3r-8}}\,,
\end{equation}
where $x$$(=m/T)$ is a dimensionless quantity, $l$ denotes the total angular momentum quantum number, representing the nature of the partial wave. For example, $l=0$ and $l=1$ describe $s-$wave and $d-$wave cross-section, respectively.  Note that this cross-section is valid for spin-0 DM particles. In the following, we consider scalar dark matter for our analysis, focusing on the $s-$wave DM annihilation process with $r=3,4$ and place the upper bound on cross-section using Eq.~(\ref{Evo_4}). The maximum allowed $s-$wave cross-sections for $3\rightarrow2$ and $4\rightarrow2$ process are given by
\begin{align}\label{max_32}
\langle\sigma_{3\rightarrow2} v^2\rangle_\text{max}^{s-\text{wave}} &= \frac{8\sqrt{2}\, (\pi x)^2}{g\, m^5}\,,
\\\label{max_42}
\langle\sigma_{4\rightarrow2} v^3\rangle_\text{max}^{s-\text{wave}} &= \frac{32\, (\pi x)^\frac{7}{2}}{g^2\, m^8}.\
\end{align}
The thermally averaged maximum cross-section is calculated at freeze-out to compare it with the cross-section yielding the correct order of DM  relic density today. Now, it is evident that one needs to know the value of $x$ at freeze-out,  $x_{\text{fo}}(=m/T_{\text{fo}})$ or $T_{\text{fo}}$ to calculate the maximum cross-section.   
Our analysis focuses solely on the $s-$wave annihilation process, which implies that the cross-section is temperature or $x$ independent quantity. Next, we discuss the DM freeze-out during different background cosmologies.

\section{Freeze-out during radiation dominated era}\label{sec4}
This section discusses the standard picture of freeze-out, where the dark matter freeze-out takes place in a radiation-dominated era. One can rewrite Eq.~(\ref{Evo_1}) in a convenient form by defining comoving dark matter abundance $Y(T)=n(T)/s(T)$, provided the entropy density is conserved during and after the freeze-out, as
\begin{equation}\label{RD1}
\frac{dY}{dx} = - \frac{ (\Delta n) \svk\, s^{r-1}}{x\, H} \left[Y^r - Y_\text{eq}^{r-2} Y^2\right],
\end{equation}
where $s$ is the entropy density and $Y_{\text{eq}}(=n_{\text{eq}}/s)$ denotes the equilibrium comoving yeild. Here, $H$ describes the Hubble rate for a radiation-dominated background. In this case, one can obtain the freeze-out temperature by solving Eq.~(\ref{Evo_5}) as
\begin{align}\label{RD2}
x_{\text{fo}}= \frac12\, \frac{3r-7}{r-1}\, \mathcal{W}\left[2\, \frac{r-1}{3r-7} \left(\frac{2^{3r-4}\, g^{2-2r}\, \gs\, \pi^{3r-1}}{45} \frac{m^{10-6r}}{M_P^2\, \svk^2}\right)^\frac{1}{7-3r}\right].
\end{align}
where $\mathcal{W}$ denotes the Lambert $\mathcal{W}$ function\footnote{If $z$ is real, for $1/e\le z < 0$, there exits two possible real values of $\mathcal{W}(z)$. One is branch $0$, denoted by $\mathcal{W}_0(z)$ (also called principle branch), provided $\mathcal{W}(z)\ge -1$. Another one is branch $-1$, symbolized by $\mathcal{W}_{-1}(z)$ satisfying $\mathcal{W}(z)\le -1$.}, but for $r>2$, the function is $\mathcal{W}_{0}$, branch $0$ of the Lambert $\mathcal{W}$ function. Now, the cross-section can be estimated analytically by solving Eq.~(\ref{RD1}) from the time of freeze-out until today (i.e., small temperature means large $x$)
\begin{align}\label{RD3}
\int_{Y_\text{fo}}^{Y_0} \frac{dY}{Y^r} \simeq \frac{Y_0^{1-r}}{1-r} \simeq - \svk \int_{\xfo}^\infty \frac{s^{r-1}}{x\, H}\, dx\,.
\end{align}
Here, $\svk$ is temperature independent (i.e., $x$ independent) since we consider $s-$ wave cross-section for DM and $Y_0$ represents yield at a very late time. Using the fact that ${Y_0}\ll {Y_{\text{fo}}}$, one can express the cross-section as
\begin{align}\label{RD4}
\svk \simeq 2^{\frac12 - r} \left(\frac{3}{\pi}\right)^{2r-3} 5^{r-\frac32}\, \frac{3r-5}{r-1}\, \frac{\sqrt{\gs}}{\gss^{r-1}}\, \frac{\xfo^{3r-5}}{M_P\, m^{2r-4}\, (m\, Y_0)^{r-1}}\,.
\end{align}
To satisfy the whole observed relic abundance of dark matter $\Omega h^2\simeq0.12$~\cite{Planck:2018vyg, Drees:2018hzm}, it is needed that
\begin{align}\label{RD5}
mY_0=\Omega h^2\frac{1}{s_0}\frac{\rho_c}{h^2}\simeq 4.3\times 10^{-10}~\text{GeV}.
\end{align}
where $s_0(\simeq 2.69\times10^3~\text{cm}^{-3})$ and $\rho_c(\simeq 1.05\times 10^{-5} h^2 ~\text{GeV}/\text{cm}^3)$ are the present day entropy density and the critical energy density, respectively~\cite{ParticleDataGroup:2022pth}. In the succeeding section, we describe the DM production during the epoch of reheating.

\section{Freeze-out during reheating}\label{sec5}
In this case, the energy budget of the universe is dominated by a component $\phi$, oscillating in a generic potential ($V(\phi)\propto\phi^k$), whose energy density scales as $\rho_\phi(a)\propto a(t)^{-3(1+\omega)}$ with $\omega={(k-2)}/{(k+2)}$. Here, it is clear that the energy density of $\phi$ redshifts like non-relativistic matter, $\rho_\phi(a)\propto a^{-3}$ for quadratic potential ($k=2$ case). It has to decay at least before the onset of BBN. So, the SM entropy is not conserved since the inflaton decays to the SM particles. Therefore, one can rewrite Eq.~(\ref{Evo_1}) in a more appropriate form as~\footnote{It is important to mention that inflaton decay to pair of dark matters can also contribute to dark matter yield. However, one can safely neglect this contribution as long as the branching ratio of the inflaton decay into dark matters $\le 10^{-4}m/(100~\text{GeV})$~\cite{Arias:2019uol, Drees:2017iod}.}
\begin{align}\label{LTR_1}
\frac{dN}{da}=-\frac{\langle\sigma_{r\rightarrow 2} v^{r-1}\rangle}{a^{3r-2}H}[N^r-N_{\text{eq}}^{r-2}N^2],
\end{align}
with $N(\equiv na^3)$ being the comoving number density and  $N_{\text{eq}}(\equiv n_{\text{eq}}a^3)$ refers to the equilibrium comoving number density of the DM. Now, using Eq.~(\ref{Inf_Reh2}), the comoving number density after the end of reheating can be estimated analytically by solving Eq.~(\ref{LTR_1}) 
\begin{align}\label{LTR_2}
N(a_{\text{rh}})=\bigg({\frac{(3r-3)(k+2)-3k}{(k+2)(r-1)}}\bigg)^{\frac{1}{r-1}}a_{\text{rh}}^3\bigg(\frac{H_{\text{rh}}}{\langle\sigma_{r\rightarrow 2} v^{r-1}\rangle}\bigg)^{\frac{1}{r-1}}\bigg(\frac{a_{\text{rh}}}{a_{\text{fo}}}\bigg)^{-3+\frac{3k}{(k+2)(r-1)}},
\end{align}
where $a_{\text{rh}}$ and $a_{\text{fo}}$ denote the scale factor at the end of reheating (i.e., at $T=T_{\text{rh}}$) and freeze-out (i.e., at $T=T_{\text{fo}}$), respectively. Once the universe enters the phase of radiation domination, the SM entropy is conserved, and so, the DM yield at present is $Y_0=Y(a_{\text{rh}})(=N(a_{\text{rh}})/s(a_{\text{rh}})a_{\text{rh}}^3)$. Now, we discuss the freeze-out of DM in two distinct scenarios where the decay modes of the inflaton are fermionic and bosonic, respectively.

\subsection{Freeze-out during fermionic reheating}
In the present case, we study the DM evolution during a reheating era, where the inflaton decays into a pair of fermions. Here, the freeze-out temperature can be calculated utilizing the Eqs. (\ref{Inf_Reh5}), (\ref{Evo_3}), and (\ref{Evo_5}) 
\begin{equation}\label{eq:FR_xfo}
\xfo =
\begin{dcases}
\frac12\, \frac{3r-7}{r-1}\, \mathcal{W}\left[2\, \frac{r-1}{3r-7} \left(\frac{2^{3r-4}\, g^{2-2r}\, \gs\, \pi^{3r-1}}{45} \frac{m^{10-6r}}{M_P^2\, \svk^2}\right)^\frac{1}{7-3r}\right]\\
\hspace{10.4cm} \text{for } \xfo \gg \xrh,\\
\frac12\, \frac{b^\prime}{(r-1)(k-1)}\,\\ \mathcal{W}\left[2\, \frac{(r-1)(k-1)}{b^\prime}\,
\left(\frac{2^{b^\prime+3k+1}\, g^{-\frac{2}{3}{(b^\prime+4k)}}\, \gs^{k-1}\, \pi^{b^\prime+6k-2}}{45^{k-1}} \frac{\xrh^4\, m^{-4-4k-2b^\prime}}{(M_P\, \svk)^{2k-2}}\right)^{-\frac{1}{b^\prime}}\right]\\
\hspace{10.4cm} \text{for } \xfo \ll \xrh,
\end{dcases}
\end{equation}
with $b^\prime=3kr-3r-7k+3$. Note that, for (${(r-1)(k-1)}/{b^\prime}<0$), the function $\mathcal{W}$ corresponds to $\mathcal{W}_{-1}$ otherwise $\mathcal{W}_{0}$. Now, the Boltzmann Eq.~({\ref{LTR_1}}) can be integrated analytically to derive the cross-section as 
\begin{align}\nonumber
\int_{N_\text{fo}}^{N_0}\frac{dN}{N^r} &\simeq \frac{N_0^{1-r}}{1-r} \simeq - \svk \left[\int_{\afo}^{\arh} \frac{da}{a^{3r-2}\, H} + \int_{\arh}^\infty \frac{da}{a^{3r-2}\, H}\right]\\
&\simeq - \svk \bigg[\frac{k+2}{3k-(3r-3)(k+2)}\frac{a_{\text{rh}}^{3-3r}}{H_{\text{rh}}}\bigg(1-\Big(\frac{a_{\text{fo}}}{a_{\text{rh}}}\Big)^{\frac{3k}{k+2}-3r+3}\bigg)+\frac{1}{3r-5}\frac{a_{\text{rh}}^{3-3r}}{H_{\text{rh}}}\bigg],
\label{eq:FR_BEq}
\end{align}
with $N_\text{fo} \equiv N(\afo)$ and $N_0$ refers to the asymptotic value of $N(a)$ at large values of $a$, much later time after freeze-out. The integral has been divided into two parts to highlight the two regimes of $H$ in Eq.~(\ref{Inf_Reh5}) since the universe evolves differently during and after reheating. Using the fact that $N_\text{fo} \gg N_0$,
the thermally-averaged cross-section is, therefore,
\begin{equation}\label{eq:FR_cross}
\svk \simeq \frac{5^{r-\frac32} 9^{r-1}}{2^{r-\frac12} \pi^{2r-3}} \frac{\tilde c (3r-5)}{(r-1)(k+2)} \frac{\sqrt{\gs}}{\gss^{r-1}} \frac{\xrh^{3r-5}}{\frac{2(4-k)}{k+2} + 2(3r-5) \left(\frac{\xrh}{\xfo}\right)^\frac{\tilde c}{k-1}} \frac{m^{4-2r}}{M_P\, (m\, Y_0)^{r-1}}\,,
\end{equation}
with $\tilde c=2(r-1)(k+2)-2k$.

\subsection{Freeze-out during bosonic reheating}
Alternatively, the inflaton could have decayed into two bosons. Then, one can estimate the freeze-out temperature with the help of Eqs. (\ref{Inf_Reh7}), (\ref{Evo_3}) and, (\ref{Evo_5}) as
\begin{equation} \label{eq:BR_xfo}
\xfo =
\begin{dcases}
\frac12\, \frac{3r-7}{r-1}\, \mathcal{W}\left[2\, \frac{r-1}{3r-7} \left(\frac{2^{3r-4}\, g^{2-2r}\, \gs\, \pi^{3r-1}}{45} \frac{m^{10-6r}}{M_P^2\, \svk^2}\right)^\frac{1}{7-3r}\right]\\
\hspace{10.4cm} \text{for } \xfo \gg \xrh,\\
\frac12\, \frac{3r-4k-3}{r-1}\,\\ \mathcal{W}\left[2\, \frac{r-1}{3r-4k-3}\,
\left(\frac{2^{3r-4}\, g^{2-2r}\, \gs\, \pi^{3r-1}}{45} \frac{\xrh^{4k-4}\, m^{10-6r}}{M_P^2\, \svk^2}\right)^\frac{1}{4k+3-3r}\right]\\
\hspace{10.4cm} \text{for } \xfo \ll \xrh,
\end{dcases}
\end{equation}
Here, for ($(r-1)/(3r-4k-3)<0$), the function $\mathcal{W}$ corresponds to $\mathcal{W}_{-1}$ otherwise $\mathcal{W}_{0}$. Using Eqs.~(\ref{eq:FR_BEq}) and (\ref{Inf_Reh5}) and considering $N_\text{fo} \gg N_0$, one can obtain the analytical expression of thermally averaged cross-section as
\begin{equation} \label{eq:BR_cross}
\svk \simeq \frac{5^{r-\frac32} 9^{r-1}}{2^{r-\frac12} \pi^{2r-3}} \frac{\tilde c (3r-5)}{(r-1)(k+2)} \frac{\sqrt{\gs}}{\gss^{r-1}} \frac{\xrh^{3r-5}}{\frac{2(4-k)}{k+2} + 2(3r-5) \left(\frac{\xrh}{\xfo}\right)^{\tilde c}}\frac{m^{4-2r}}{M_P\, (m\, Y_0)^{r-1}}\,.
\end{equation}
Next, we will show our results and discuss elaborately the features of the parameter space of DM.

\section{Results}\label{sec6}
In this section, we discuss the results of the DM freeze-out at different eras, like the radiation-dominated era (i.e., $T_{\text{fo}}\ll T_{\text{rh}}$) and the fermionic and bosonic reheating epoch (i.e., $T_{\text{fo}}\ge T_{\text{rh}}$). In addition, we show various constraints on the parameter space, such as BBN, no chemical equilibrium, and relativistic freeze-out constraint, along with the unitarity bound.
\begin{figure}[htb!]
	\centering
	\includegraphics[height=4.2cm,width=5.15cm]{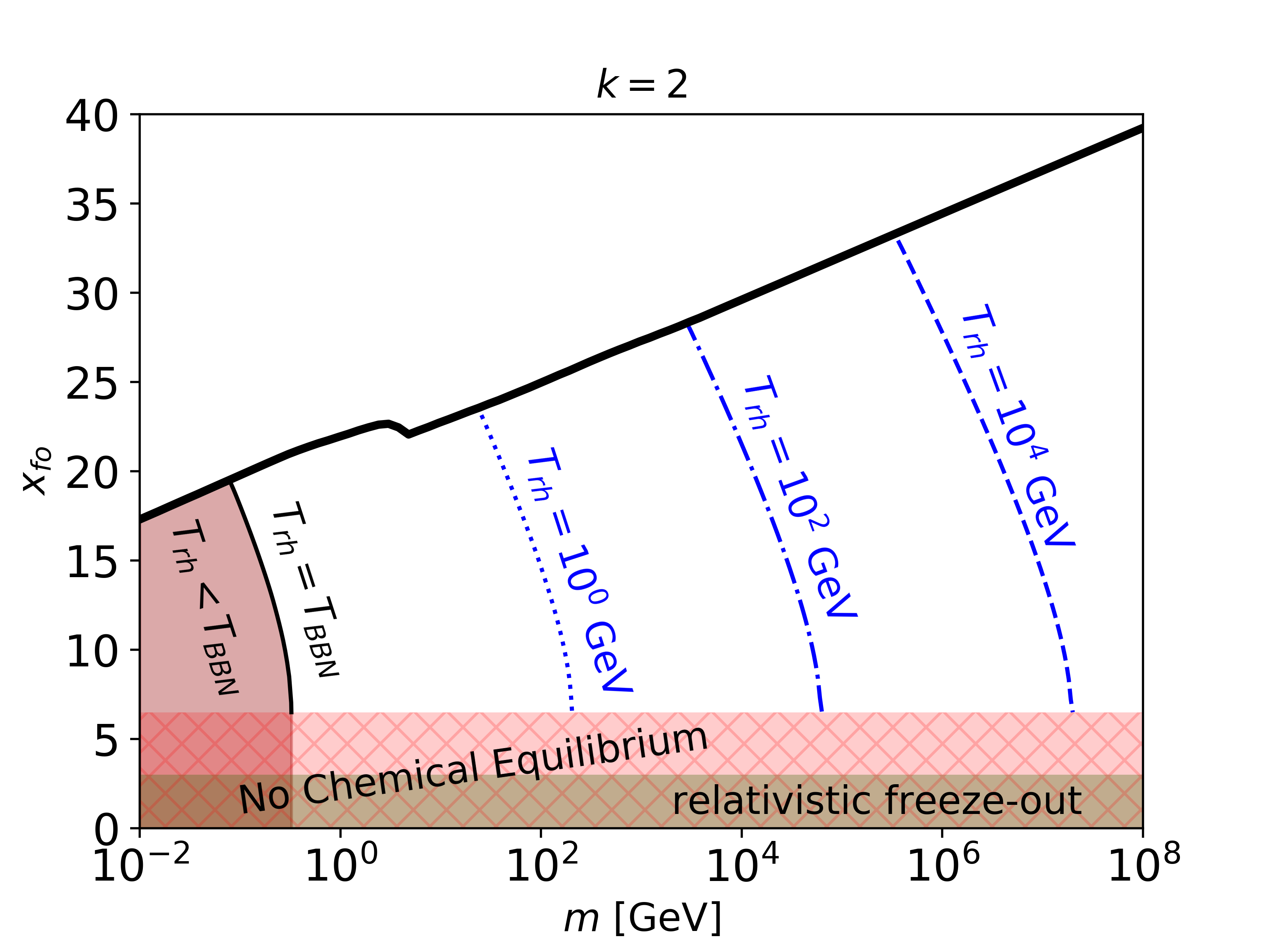}\hfil
	\includegraphics[height=4.2cm,width=5.15cm]{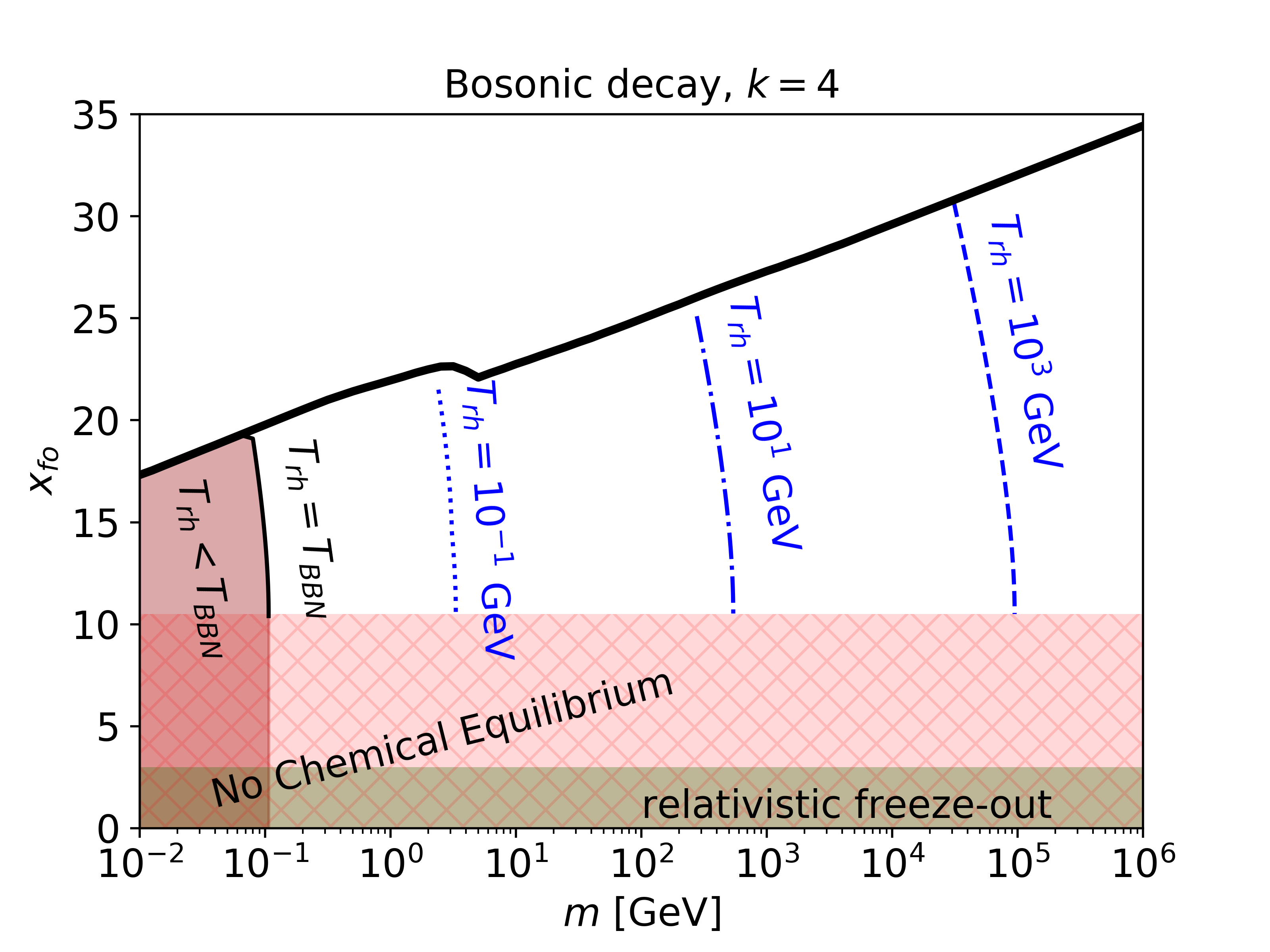}\hfill
	\includegraphics[height=4.2cm,width=5.15cm]{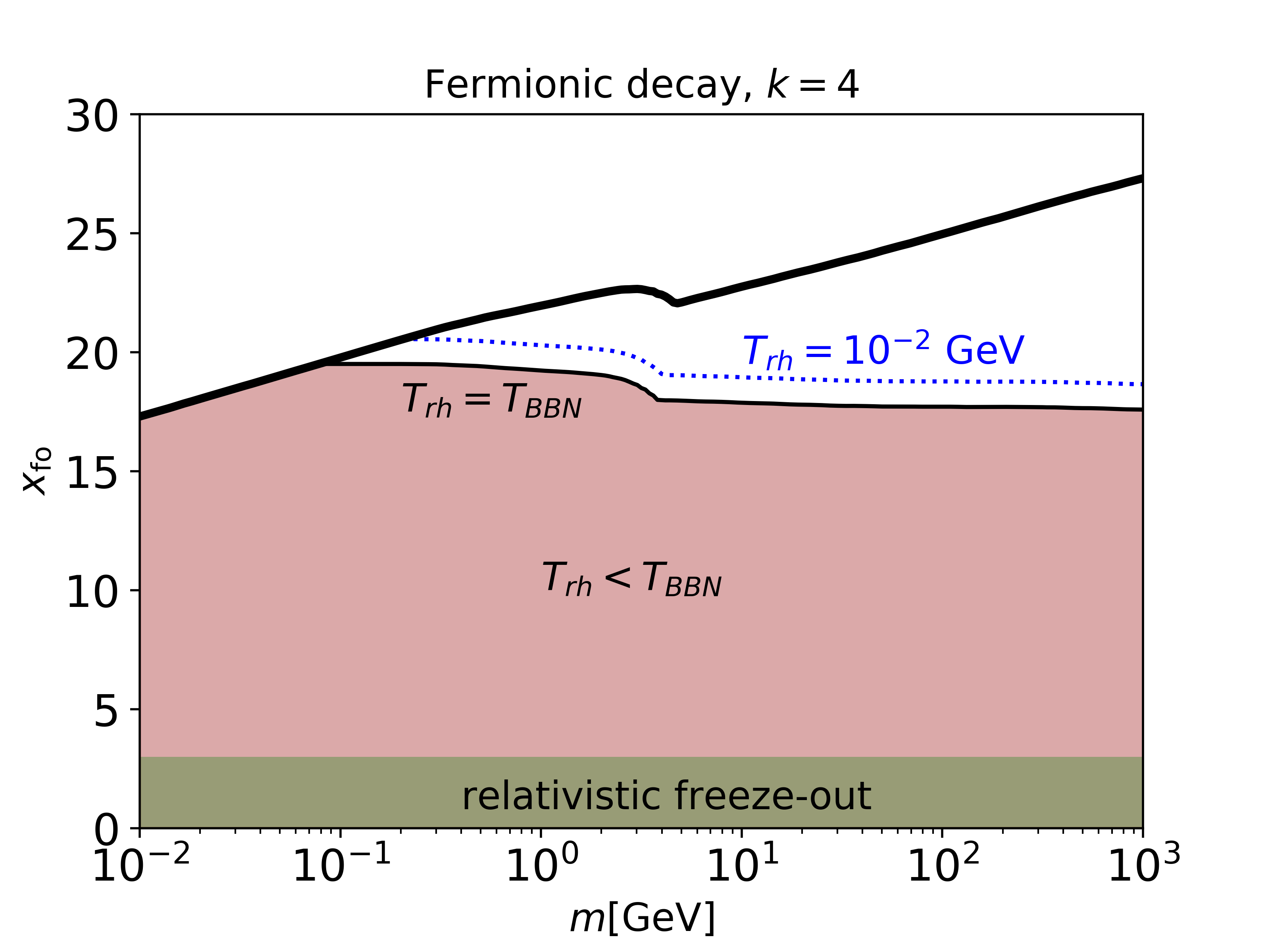}
	\caption{Plots show the freeze-out temperature $T_{\text{fo}}(=m/x_{\text{fo}})$ needed to match the observed DM abundance as a function of DM mass for radiation-dominated background (thick black line) and different reheating temperatures $(T_{\text{rh}})$. The left plot describes the freeze-out for the reheating scenario with quadratic inflaton potential (i.e., $k=2$ case), where the middle and right plots refer to the freeze-out during fermionic and bosonic reheating, respectively, provided $k=4$. The no chemical equilibrium constraint disallows the red-shaded hatched region. The brown shaded region denoted by $T_{\text{rh}}<T_{\text{BBN}}$ is disfavoured by BBN constraint. The relativistic freeze-out constraint rules out the region $x_{\text{fo}}\le 3$.}
	\label{fig:xf_32}
\end{figure}
Figure~\ref{fig:xf_32} shows the freeze-out temperature required to reproduce the observed relic abundance for DM as a function of DM mass, considering the freeze-out during the radiation-dominated era and different reheating eras. The left plot demonstrates the freeze-out during the reheating scenario with $k=2$. In contrast, the middle and right plots display the same for reheating pictures with $k=4$ for the bosonic and fermionic cases, respectively. The thick black solid line corresponds to standard freeze-out in the radiation epoch. The left plot displays the freeze-out temperature for radiation, $T_{\text{rh}}=T_{\text{BBN}}$ (black solid line), $T_{\text{rh}}=1$ GeV (blue dotted line), $T_{\text{rh}}=10^2$ GeV (blue dot-dashed line), $T_{\text{rh}}=10^4$ GeV (blue dashed line). It is essential to mention that for the bosonic case (middle plot), the blue dotted, dot-dashed, and dashed line dictate the reheating temperature ($T_{\text{rh}}$) $0.1$ GeV, $10$ GeV and, $10^3$ GeV, respectively. 

During reheating, entropy production takes place, so overproduction of DM at freeze-out is needed to compensate for the entropy dilution effect, which leads to earlier freeze-out, therefore a smaller value of $x_{\text{fo}}$. However, $x_\text{fo}$ cannot be very small since the chemical equilibrium of $3\rightarrow 2$ process demands $x_{\text{fo}}\ge 6.5$, which is obvious from the left plot of Figure~\ref{fig:xf_32}. The red-shaded hatched area displays the no chemical equilibrium region where DM never achieves chemical equilibrium. BBN constraint $(T_{\text{rh}}\ge T_{\text{BBN}})$ disallows a few portions of parameter space of each plot, shown by the brown shaded region. Now, the variation of $x_{\text{fo}}$ with DM mass for $r\rightarrow2$ process for a fixed $T_{\text{rh}}$ can be understood by looking at the following scaling, which can be deduced by using the freeze-out criterion (Eqs.~(\ref{Evo_3})) and the expressions of number density, cross-sections and, Hubble rates as
\begin{align}\label{rs1}\nonumber
&x_{\text{fo}}\propto \log (m) ~~~~~~~~~~~~~~~~~~~~~~\text{radiation domination}\\
&x_{\text{fo}}\propto \frac{2(k-4)}{k-1}\log (m) ~~~~~~~~~~\text{fermionic reheating}\\\nonumber
&x_{\text{fo}}\propto -2k\log (m) ~~~~~~~~~~~~~~~~\text{bosonic reheating}
\end{align}
It is important to notice that the mass scaling of $x_\text{fo}$ is independent of $r$ irrespective of the background cosmologies, i.e., it does not depend on the number of initial state particles. $x_{\text{fo}}$ varies logarithmically for radiation-dominated picture. For $k=2$ case, $x_\text{fo}$ falls as $\log(m^{-4})$, independent of inflaton decay modes for a fixed reheating temperature, which is obvious from the left plot of Figure~\ref{fig:xf_32}. Note that, $x_\text{fo}$ ($\propto\log (m^{-8})$) falls more rapidly for bosonic reheating case with $k=4$ compared to $k=2$ case. That is why the contours of $x_{\text{fo}}$ are steeper for $k=4$ bosonic reheating picture than $k=2$ case. Interestingly, $x_\text{fo}$ is independent of DM mass for fermionic reheating with $k=4$ and $x_{\text{fo}}$ contours are parallel to the DM mass axis, which is clear from the right plot of Figure~\ref{fig:xf_32}. The bump in the relic density satisfied contours is due to the sudden change in the number of relativistic degrees of freedom near the QCD phase transition ($T\sim 0.1$ GeV). 

\begin{figure}[htb!]
	\centering
	\includegraphics[height=4.2cm,width=5.15cm]{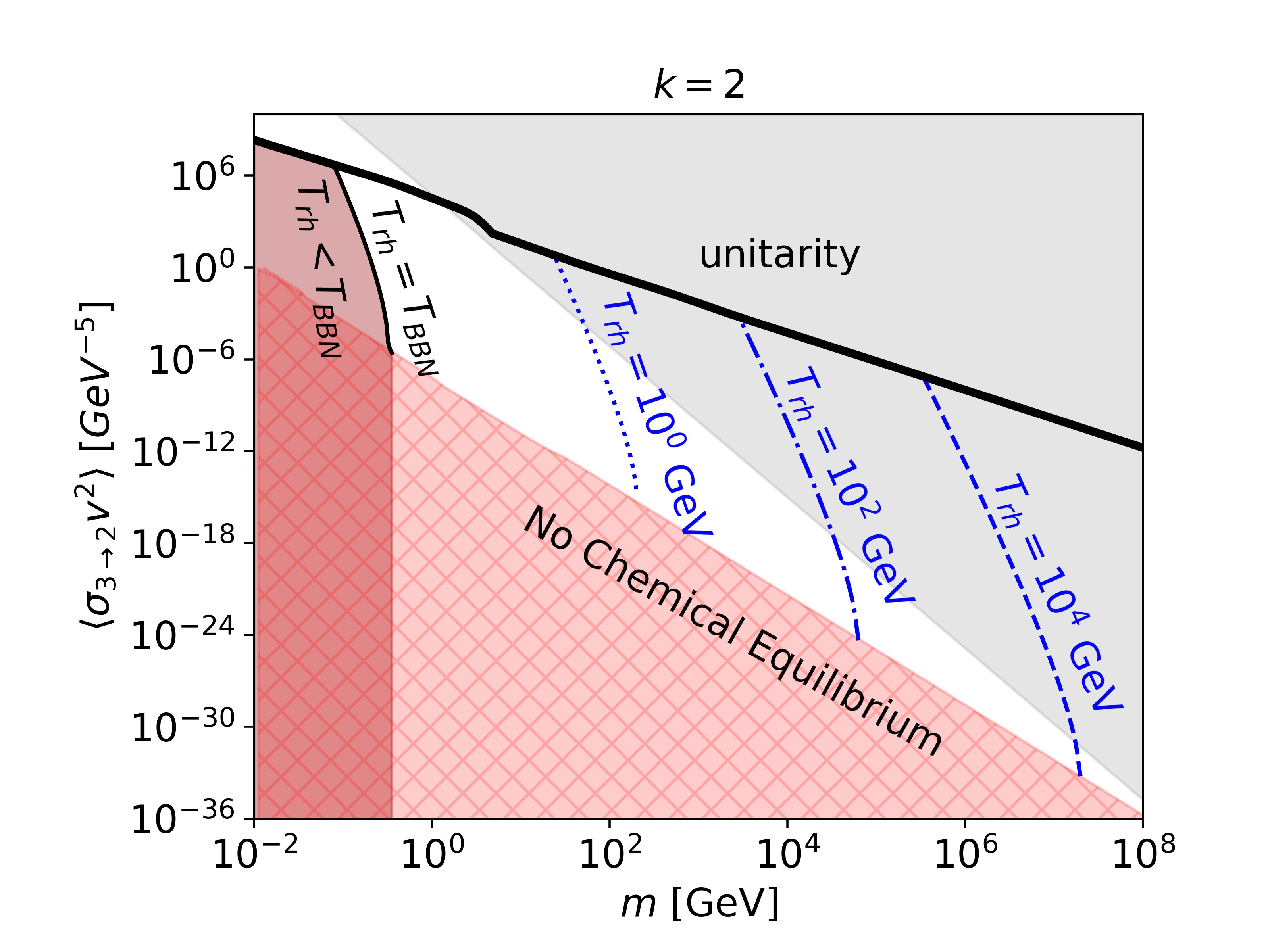}\hfill
	\includegraphics[height=4.2cm,width=5.115cm]{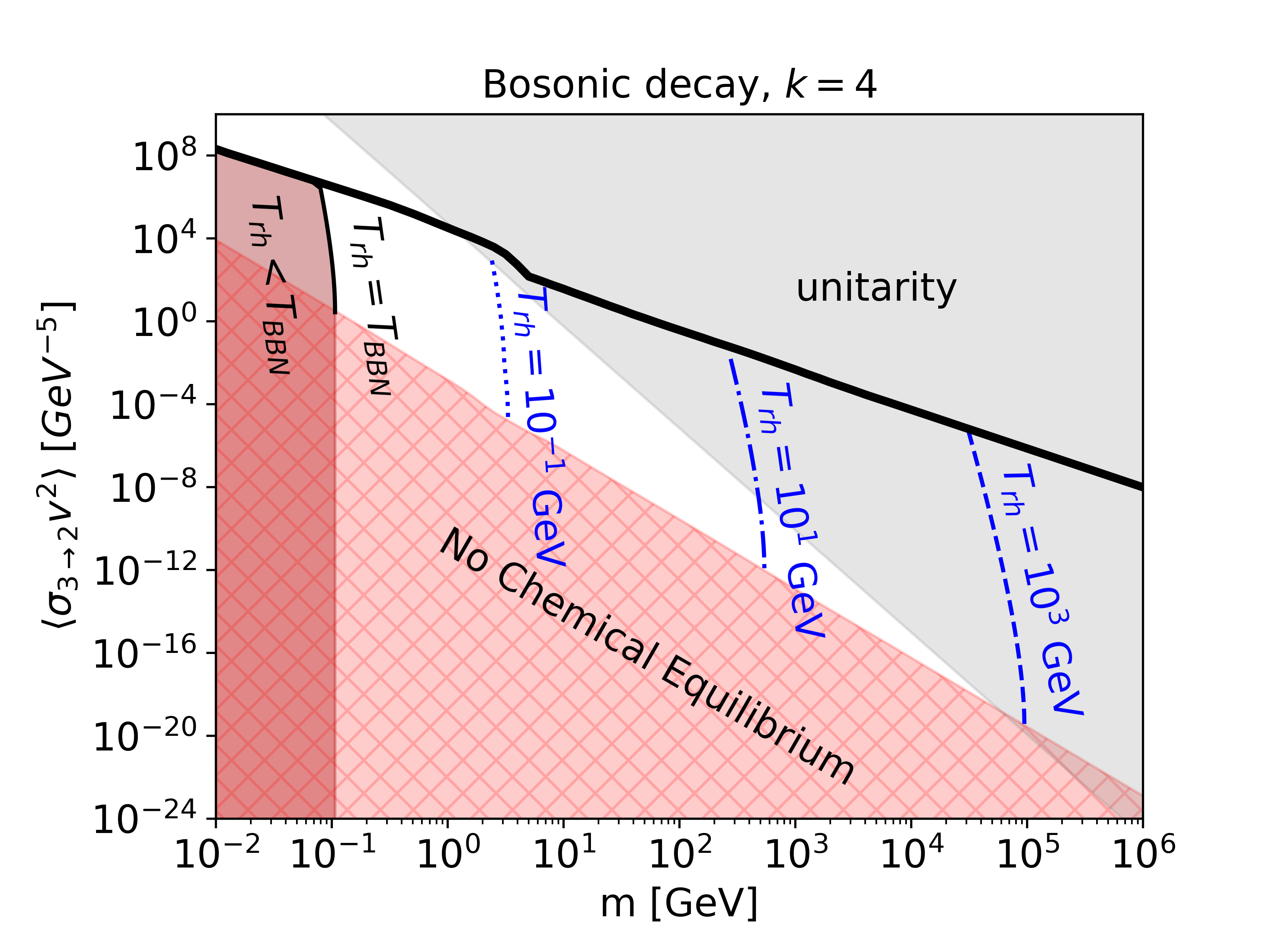}\hfill
	\includegraphics[height=4.2cm,width=5.15cm]{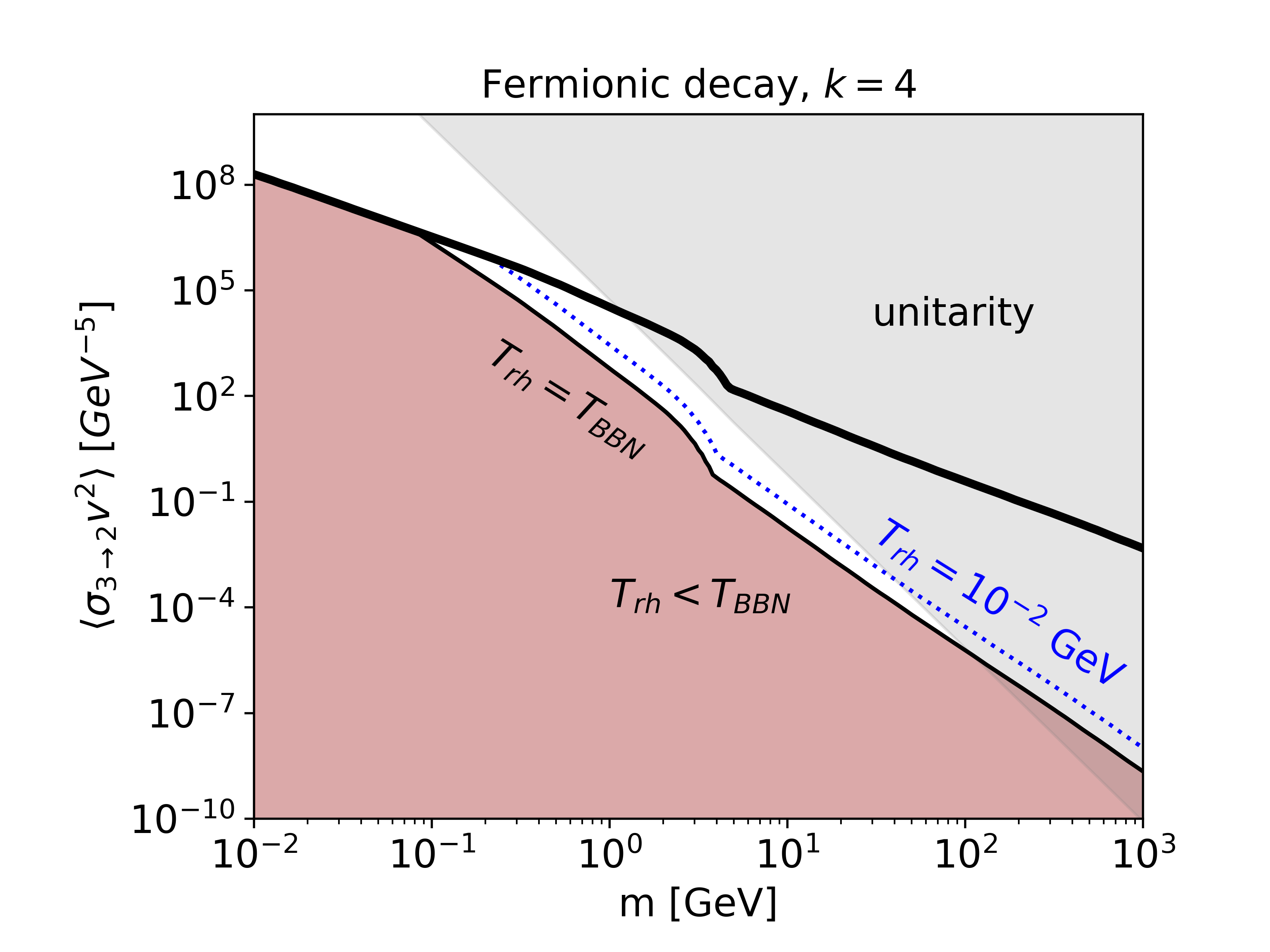}
	\caption{Cross-section that satisfies the observed dark matter abundance for $3\rightarrow 2$ annihilation of SIMP DM in case of reheating with $k=2$ (left plot) and $k=4$ (middle and right plots) for different reheating temperatures. The middle and right plots display the impact of bosonic and fermionic reheating on cross-section and dark matter mass planes. The black thick solid line, blue dotted, blue dotdashed and blue dashed line are the relic satisfied cross-section for reheating temperatures $T_{\text{BBN}}$, $1$ GeV ($1$ GeV), $10^2$ GeV ($10$ GeV) and $10^4$ GeV ($1$ GeV) in case of $k=2$ reheating ($k=4$ bosonic reheating), respectively. The brown region denoted by $T_{\text{rh}}<T_{\text{BBN}}$ is ruled by the BBN constraints. The grey and red-shaded hatched regions are disallowed by unitarity and no chemical equilibrium constraints.}
	\label{fig:cross_32}
\end{figure}

Equivalent information as Figure~\ref{fig:xf_32} is also provided by Figure~\ref{fig:cross_32} but in $[ m\,,\langle\sigma_{3\rightarrow 2} v^2\rangle]$ plane. Cross-sections needed to fit the observed dark matter abundance for $k=2$, $k=4$ bosonic, and fermionic reheating have been shown by the contours for different $T_{\text{rh}}$ in Figure~\ref{fig:cross_32}. The gray shaded region is excluded by the requirement of $S$-matrix unitarity following Eq.~(\ref{max_32}). Additionally, BBN and no chemical equilibrium constraints ruled out brown-shaded and red-shaded hatched regions, respectively. We have already discussed that an early freeze-out occurs during reheating, which demands a smaller cross-section than the standard case. Hence, the reheating picture allows us to explore smaller cross-sections, obvious from Figure~\ref{fig:cross_32}. For instance, for DM mass $m\sim 200~\text{GeV}$, required cross-section is $\langle\sigma_{3\rightarrow 2} v^2\rangle\sim \mathcal{O}(10^{-1})~\text{GeV}^{-5}$ in standard radiation-dominated case where the needed cross-section can be as small as  $\langle\sigma_{3\rightarrow 2} v^2\rangle\sim \mathcal{O}(10^{-15})~\text{GeV}^{-5}$ for $T_{\text{rh}}=1~\text{GeV}$ and $k=2$ scenario. Larger $T_{\text{rh}}$ allows us to explore even smaller cross-sections for heavier DM. Each contour of fixed $T_{\text{rh}}$ reflects the fact that $\langle\sigma_{3\rightarrow 2} v^2\rangle$ decreases as the DM mass increases. This is because the freeze-out occurs earlier (i.e., for larger $T_{\text{fo}}$) for heavier SIMP, which implies a smaller $\langle\sigma_{3\rightarrow 2} v^2\rangle$. Note that, for a fixed $T_{\text{rh}}$, a range of DM mass is allowed for which one can satisfy the relic density constraints. Although the reheating scenario provides us the handle to explore smaller cross-sections, the unitarity constraint is so stringent that we end up with very small viable parameter space after considering the no chemical equilibrium constraint. We want to highlight that the parameter space shrinks with increasing $k$. The reason behind this is that the dilution effect due to entropy injection will be less prominent as the value of $k$ increases for a fixed annihilation process where the dilution factor is $(a_{\text{rh}}/a_{\text{fo}})^{-3+3k/2(k+2)}$ ((see Eq.~(\ref{LTR_2})) for $r=3$.

\begin{figure}[htb!]
	\centering
	\includegraphics[height=4.2cm,width=5.15cm]{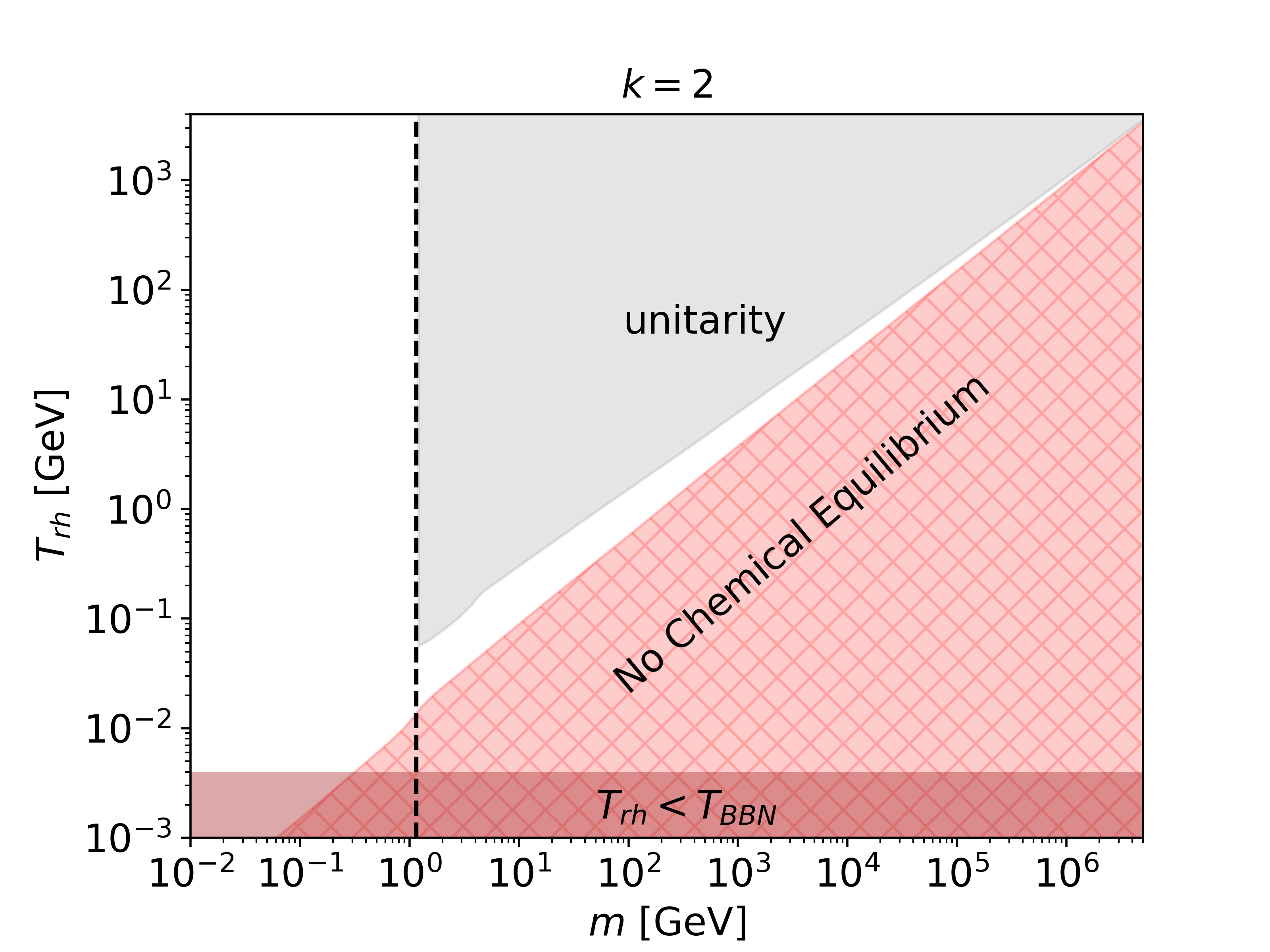}\hfill
	\includegraphics[height=4.2cm,width=5.15cm]{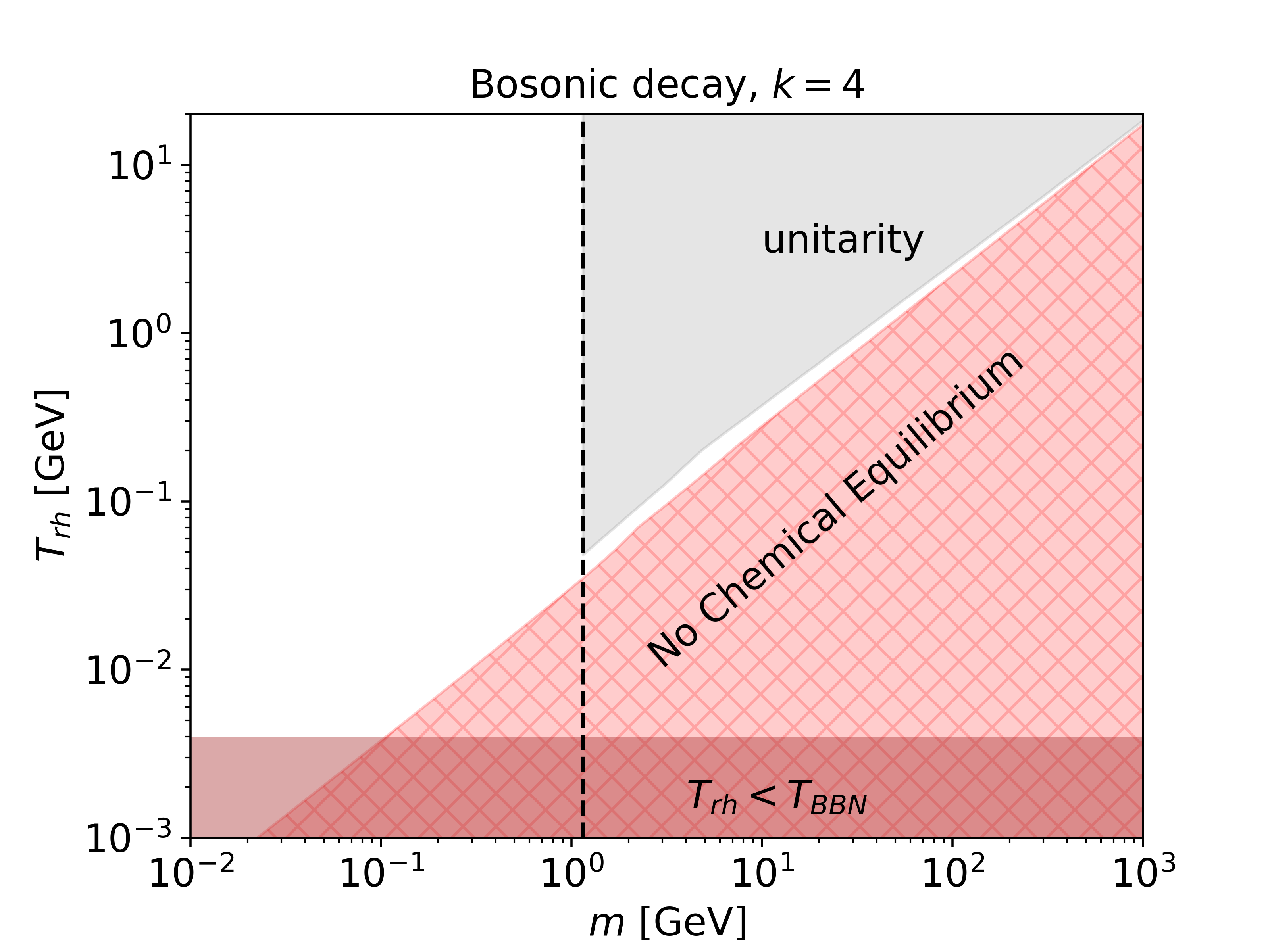}\hfill
	\includegraphics[height=4.2cm,width=5.15cm]{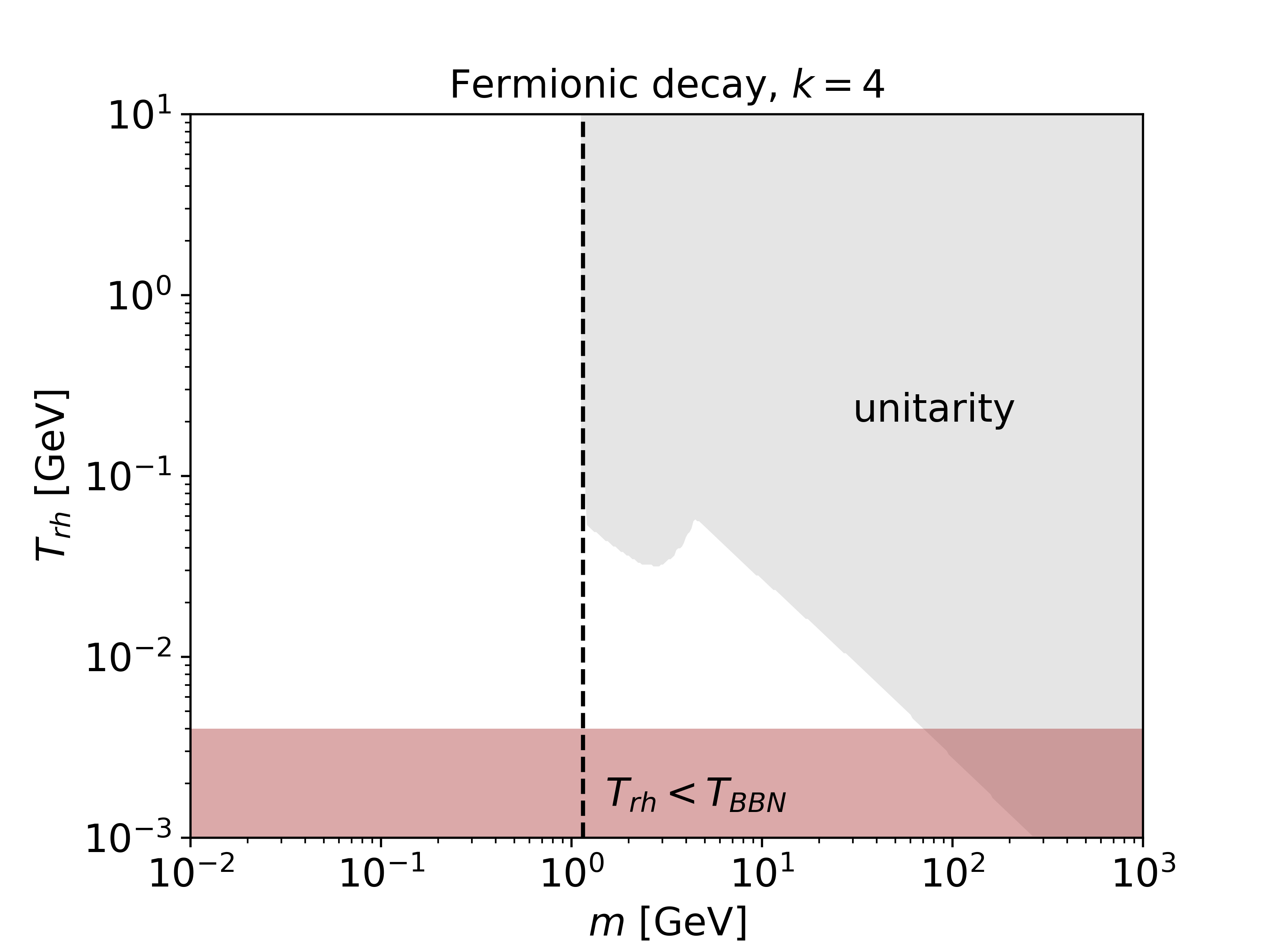}\hfill
	\caption{unitarity bound (gray shaded region) and no chemical equilibrium constraints (red shaded hatched area) in $[m,~T_{\text{rh}}]$ plane for $k=2$ reheating (left plot) and bosonic (the middle plot) and fermionic reheating (the right plot) with $k=4$. The brown band mentioned by $T_{\text{rh}}<T_{\text{BBN}}$ is disallowed by BBN constraint.}
	\label{fig:TR_m32}
\end{figure}

We have seen in Figure~\ref{fig:cross_32} that there is a viable range of DM mass for each $T_\text{rh}$, which in turn dictates that there is an allowed range of $T_\text{rh}$ for a fixed DM mass. In Figure~\ref{fig:TR_m32}, we illustrate the permissible parameter space in reheating temperature and DM mass plane where the white region is free from any constraints. The vertical black dotted line indicates DM mass around 1 GeV, which is the unitarity upper limit on DM mass, and so the region left to this line is allowed for a radiation-dominated background. However, we are mainly interested in the mass range above 1 GeV since it is ruled out in the standard radiation-dominated case. Unitarity (the gray-shaded area) confined $T_{\text{rh}}$ from above. In contrast, the necessity of chemical equilibrium (the red-shaded hatched area) bounded it from below, resulting in a strictly constrained range. Additionally, $T_{\text{rh}}$ lower than $T_{\text{BBN}}$ is in disagreement with the cosmological observation, which is depicted by the brown band. The dip in the unitarity bound in the right plot of Figure~\ref{fig:TR_m32} is the reflection of the sudden change in the effective number of relativistic degrees of freedom near $QCD$ phase transition which is also visible in freeze-out and cross-section plots. It is worth noting that while entropy injection is common in both the $\omega=0$ and $\omega=1/3$ reheating scenarios, the universe's expansion differs. The expansion is matter-like for the $\omega=0$ scenario and radiation-like for the $\omega=1/3$ scenario. Consequently, the viable parameter space is more significant for the $\omega=0$ reheating case than the $\omega=1/3$ reheating scenario.

Additionally, we have done the same analysis for SIMP dark matter where the dominant number-changing process is $4\rightarrow 2$ instead of $3\rightarrow 2$; the results and discussion can be found in the appendix~\ref{SIMP42}.

Before closing the section, we want to comment on realizing reheating potential $(\propto \phi^k)$ in the case of inflationary models where a lower bound on reheating temperature exists considering the constraints of the CMB observations. The mentioned form of the reheating potential can be reproduced for $\alpha-$attractor E- and T-model, which have the following forms~\cite{Kallosh:2013hoa, Kallosh:2013maa, Garcia:2020wiy, Becker:2023tvd}.
\begin{align}
V(\phi) &=M^4\bigg(1-e^{-\sqrt{\frac{2}{3\alpha}}\frac{\phi}{M_P}}\bigg)^{2p},~~~~~~~~~\text{E-model},\\
V(\phi)&=M^4\bigg[\tanh{\bigg(\frac{\phi}{\sqrt{6\alpha}M_P}\bigg)}\bigg]^{2p},~~~~~~\text{T-model},
\end{align}
where $M$ being a normalization energy scale and $\alpha(>0)$ and $p(>1)$ are dimensionless parameters. One can obtain the same form of potential as in Eq.~(\ref{Inf_Reh1}) for $k=2p$ and, $\lambda=\big(\frac{2}{3\alpha}\big)^{p}\big(\frac{M}{M_P}\big)^4$ in case of E-model and $\lambda=\big(\frac{1}{6\alpha}\big)^{p}\big(\frac{M}{M_P}\big)^4$ in T-model by performing the matching after expanding the potential in small field limit (i.e, $\phi<M_P$).

It is essential to mention that these models predict some particular relation among the inflaton potential parameters and their observables. These observables are the scalar to tensor ratio, spectral index, the amplitude of primordial scalar perturbation, and a number of $e-$ folds between the pivot scale $k_\star$ and the time when the inflation ends. Interstingly, the CMB experiments constrain these quantities.\cite{thecorecollaboration2011corecosmicoriginsexplorer, Matsumura:2013aja, Abazajian:2019eic}
\begin{align}\label{reh_constraint}
T_{\text{rh}}=\bigg[\bigg(\frac{43}{11g_{\star,\text{rh}}}\bigg)^{\frac{1}{3}}\frac{a_0T_0}{k_\star}H_\star e^{-N_\star}\bigg(\frac{45V_{\text{end}}}{\pi^2g_{\star,{\text{rh}}}}\bigg)^{-\frac{1}{3(1+\omega)}}\bigg]^{\frac{3(1+\omega)}{3\omega-1}}.
\end{align}

\begin{figure}[htb!]
	\centering
	\includegraphics[height=4.2cm,width=5.15cm]{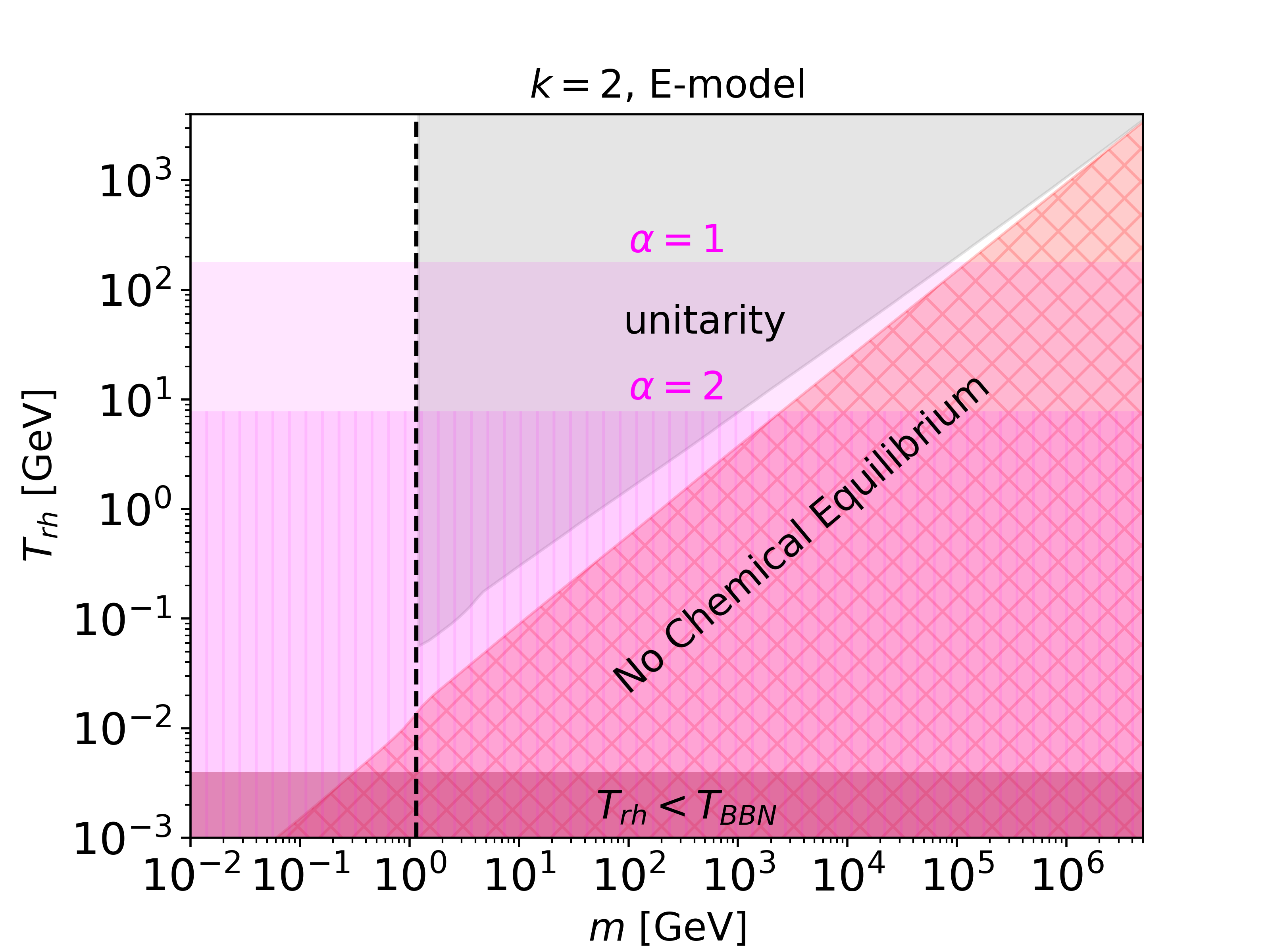}
	\includegraphics[height=4.2cm,width=5.15cm]{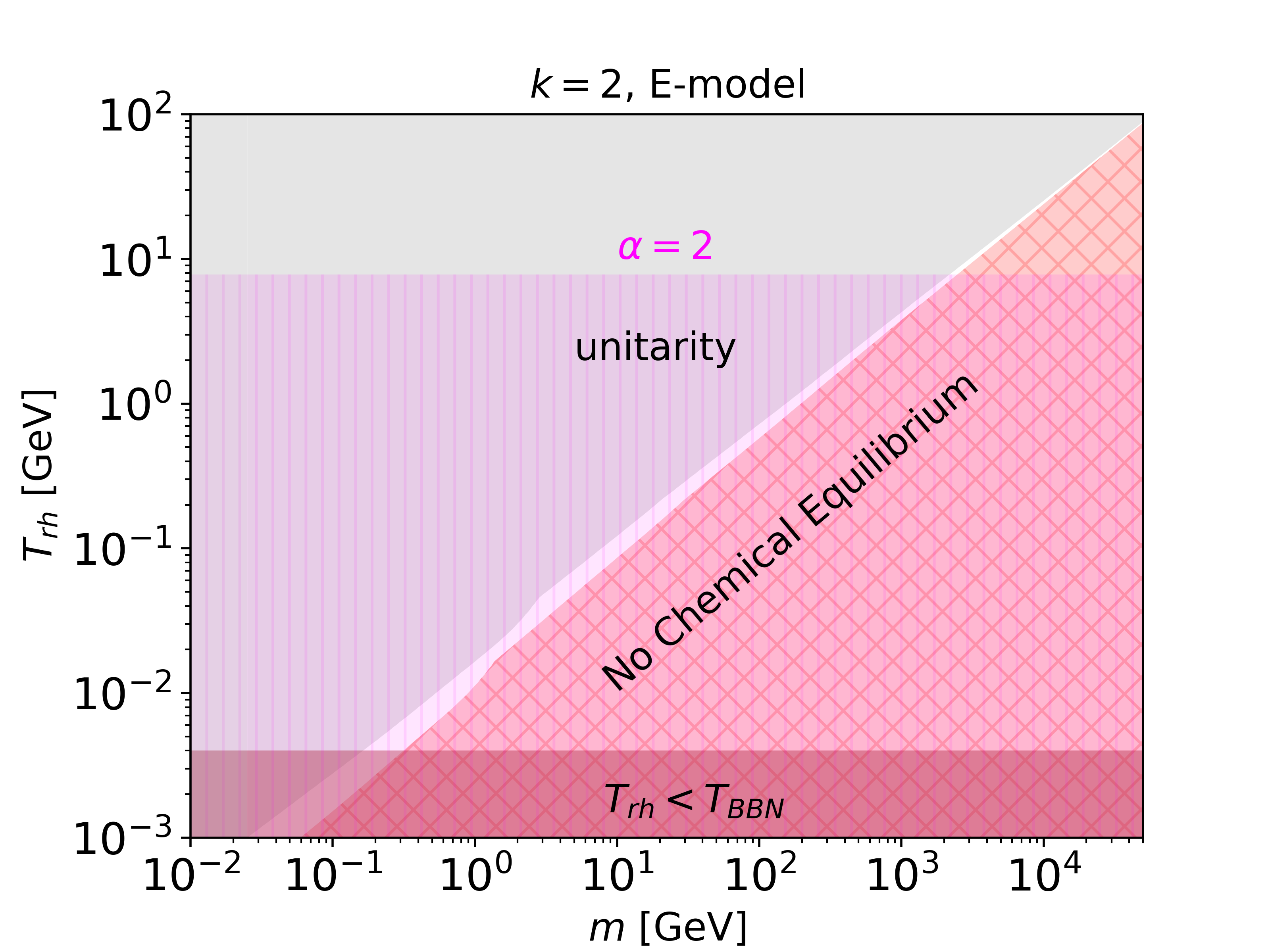}
	\caption{Shows the inflationary constraints on the reheating temperature and DM mass place for quadratic potential ($k=2$) case. Left plot displays the E-model constraint for $\alpha=1,2$ for $3\rightarrow2$ process and right plot shows the same for $\alpha=2$ for $4\rightarrow2$ process.}
	\label{fig:Inf_constraint}
\end{figure}

The dynamics of reheating affect the moment in time at which the pivot scale of CMB reenters the horizon. With this in mind, one can derive bounds on reheating temperature for different inflationary models by using the expression of $T_{\text{rh}}$ relating the inflationary observables with the reheating parameters~\cite{Dai:2014jja, Becker:2023tvd}
where $a_0$ and $T_0$ denote the scale factor and temperature at present time, $H_\star$ is the Hubble rate at $k_\star$, $V_{\text{end}}$ represents the potential at the end of inflation and $\omega$ ($\ne 1/3$) refers to the equation of state during reheating. Note that, for $\omega=\frac{1}{3}$ (i.e., for $k=4$), the Eq.~(\ref{reh_constraint}) becomes independent of $T_{\text{rh}}$ and so the prediction for $T_{\text{rh}}$ is not possible. Considering the latest constriants on the inflationary observable from Planck+BICEP/Keck+WMAP~\cite{BICEP:2021xfz}, the lower bound on E-model (T-model) for $\alpha=1$ and $\alpha=2$ are $1.8\times10^2$ ($4\times10^4$) GeV
and $7.8$ ($1.8\times10^4$) GeV respectively. We have taken these bounds on reheating temperature from the ref~\cite{Becker:2023tvd}; see for more details.

Finally, we have shown the inflationary lower bounds on $T_{\text{rh}}$ in the $T_{\text{rh}}$ and DM mass plane for $k=2$ scenario in Figure~\ref{fig:Inf_constraint}. The left (right) of Figure~\ref{fig:Inf_constraint}, displays the E-model constraint for $\alpha=1,2$ ($\alpha=2$) provided the DM number changing process is $3\rightarrow2$ ($4\rightarrow2$). The pink-shaded and pink-shaded hatched regions are disallowed by inflationary constraint for the E-model with $k=2$ for $\alpha=1$ and $\alpha=2$, respectively. It is important to mention that the allowed parameter space in $[m, T_{\text{rh}}]$ plane is ruled out totally if one considers the T-model constraints (for both the $\alpha$'s) for $3\rightarrow2$ process and same in addition to E-model constraint for the $\alpha=1$ for $4\rightarrow2$. However, the allowed parameter space of $[m, T_{\text{rh}}]$ in case of $k=4$ remains free of such inflationary constraint since $T_{\text{rh}}$ remains unconstrained.

\section{Summary and Conclusion}\label{sec7}
The collisionless cold WIMP dark matter paradigm is a widely explored dark matter picture. Despite numerous experiments conducted at various energy frontiers, no definitive evidence of WIMPs has been found yet. This motivates us to investigate alternative dark matter paradigms. The SIMP dark matter model has attracted attention recently due to its potential to alleviate apparent small-scale issues through its strong self-interactions.

The primary goal of this work is to examine the thermal freeze-out of SIMP dark matter during the period of inflationary reheating when the inflaton field oscillates coherently around the minimum of a generic potential $(\propto\phi^k)$. As the inflaton decays to SM particles during reheating, it generates entropy. To counteract the dilution effect caused by the injection of entropy, overproduction of dark matter is necessary, leading to an early freeze-out. In this case, the reheating period allows for exploring smaller dark matter cross-sections that satisfy the observed relic density compared to the standard scenario. Our investigation mainly addresses the reheating cases with \(k=2\) and \(k=4\), considering both the bosonic and fermionic decay modes of the inflaton.

The maximum cross-section allowed by $S$-matrix unitarity and the same needed to match the observed relic jointly sets an upper bound on DM mass. The upper bound on the mass of SIMP DM for the dominant DM number changing process $3\rightarrow 2$ and $4\rightarrow2$ are $1$ GeV and $7$ MeV, respectively, provided the background is radiation-dominated. Freeze-out during the reheating-dominated phase relaxes this bound by demanding smaller cross-sections than the standard radiation-dominated case. Here, the cross-section is bounded from below, requiring the chemical equilibrium of $3\rightarrow 2$ (or $4\rightarrow 2$) process of SIMP DM. In this case, the upper bound on DM mass is jointly set by the unitarity limit and the no chemical equilibrium constraint, respectively. We have found that the upper bound of $3\rightarrow2$ SIMP for $k=2$ is $10^6$ GeV, where the same for $k=4$ bosonic and fermionic reheating are $300$ and $68$ GeV, respectively. Such inflationary potential can be realized in $\alpha$ attractor E- and T- models. Cosmological observation puts a lower bound on reheating temperature for the mentioned model. Considering such bound significantly or entirely excludes our viable parameter space for the $k=2$ case.

Finally, we do not have observational evidence that the early universe was radiation-dominated before BBN. Therefore, it is crucial to consider other cosmological scenarios. In this context, we have investigated the picture of late-time reheating, which allows us to achieve thermal SIMP with masses of a few orders of GeV by relaxing the unitarity bound by requiring a smaller cross-section than the standard radiation-dominated scenario.

\acknowledgments
SS would like to thank Nicolás Bernal for many enlightening discussions. SS is supported by NPDF grant PDF/2023/002076 from the Science and Engineering Research Board (SERB), Government of India. This research of DC is supported by an initiation grant IITK/PHY/2019413 at IIT Kanpur and by a DST-SERB grant SERB/CRG/2021/007579.
\appendix
\section{{Results for $4\rightarrow 2$ SIMP}}\label{SIMP42}
{\begin{figure}[H]
		\centering
		\includegraphics[height=4.2cm,width=5.15cm]{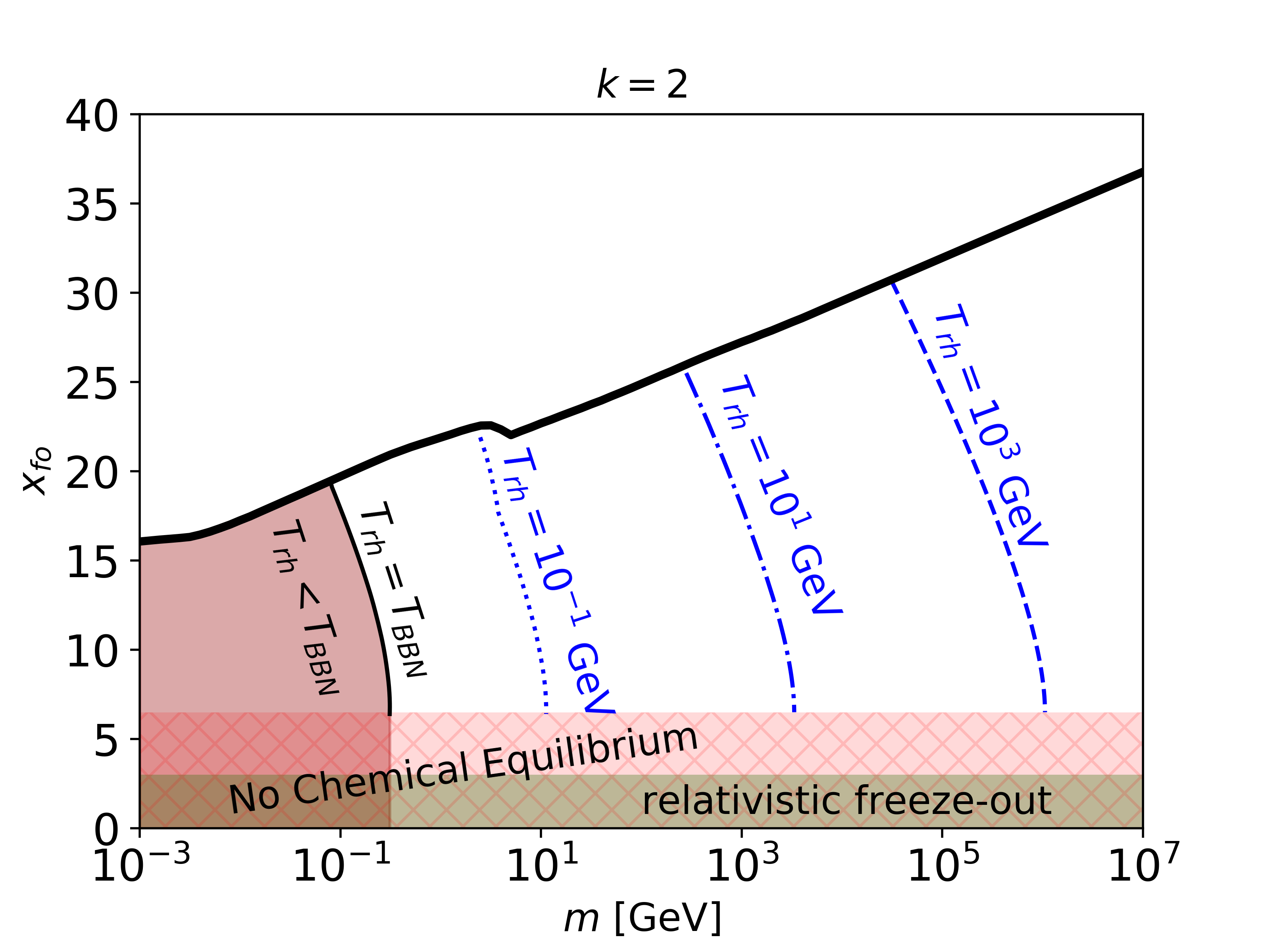}\hfill
		\includegraphics[height=4.2cm,width=5.15cm]{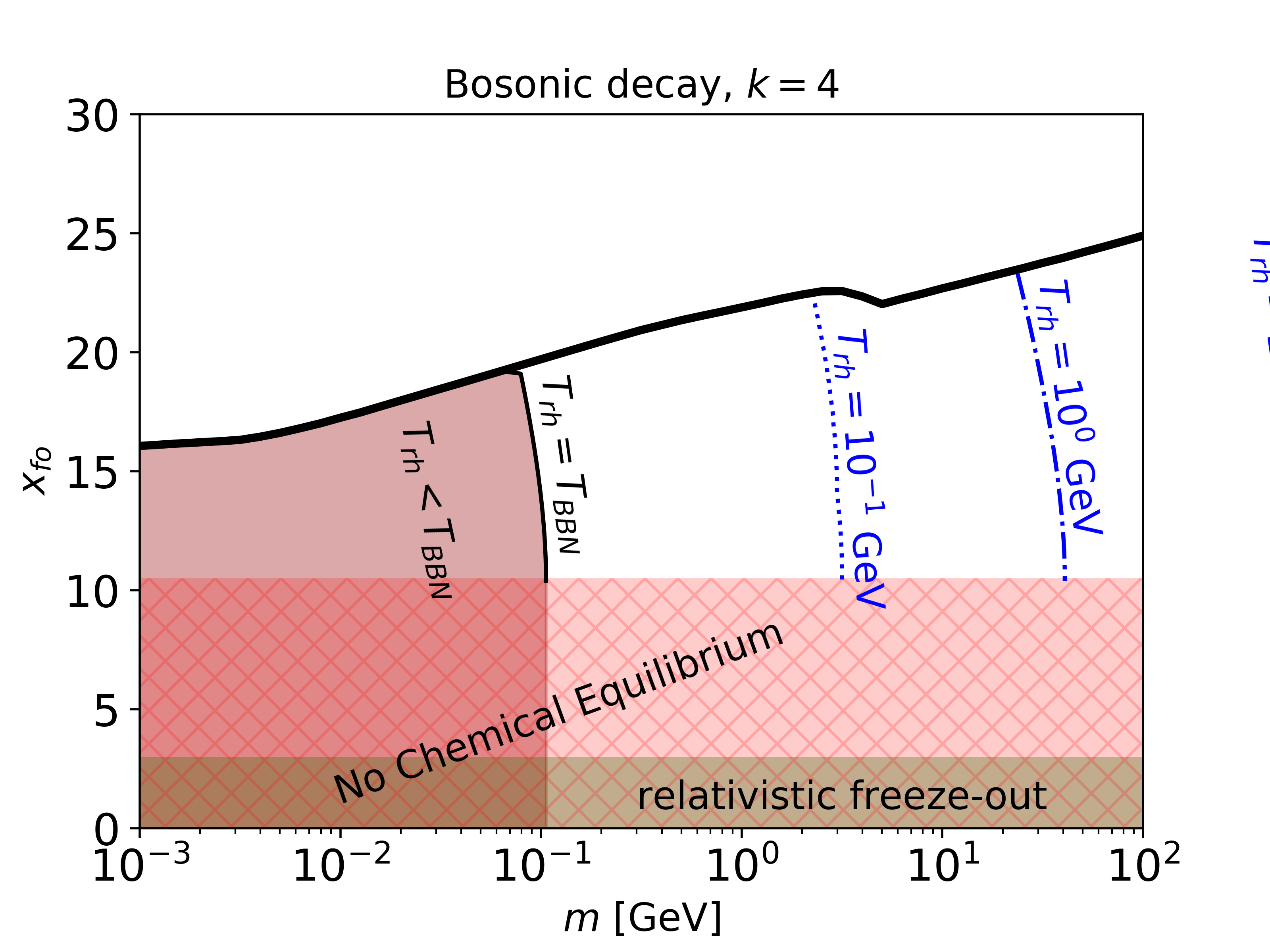}\hfill
		\includegraphics[height=4.2cm,width=5.15cm]{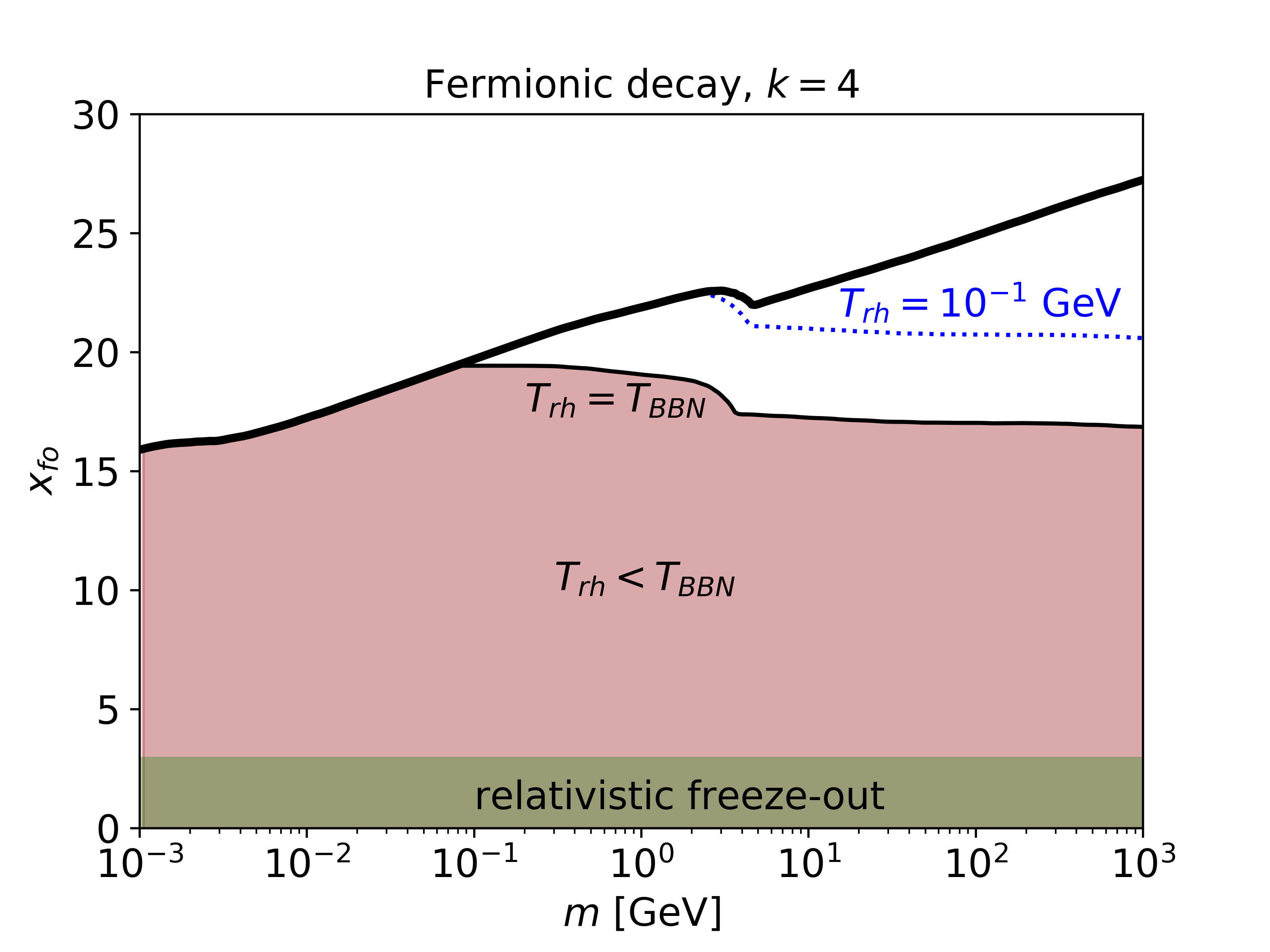}
		\caption{Same as Figure~\ref{fig:xf_32}, but for $4\rightarrow2$ process.}
		\label{fig:xf_42}
	\end{figure}
	
	
	\begin{figure}[H]
		\centering
		\includegraphics[height=4.2cm,width=5.15cm]{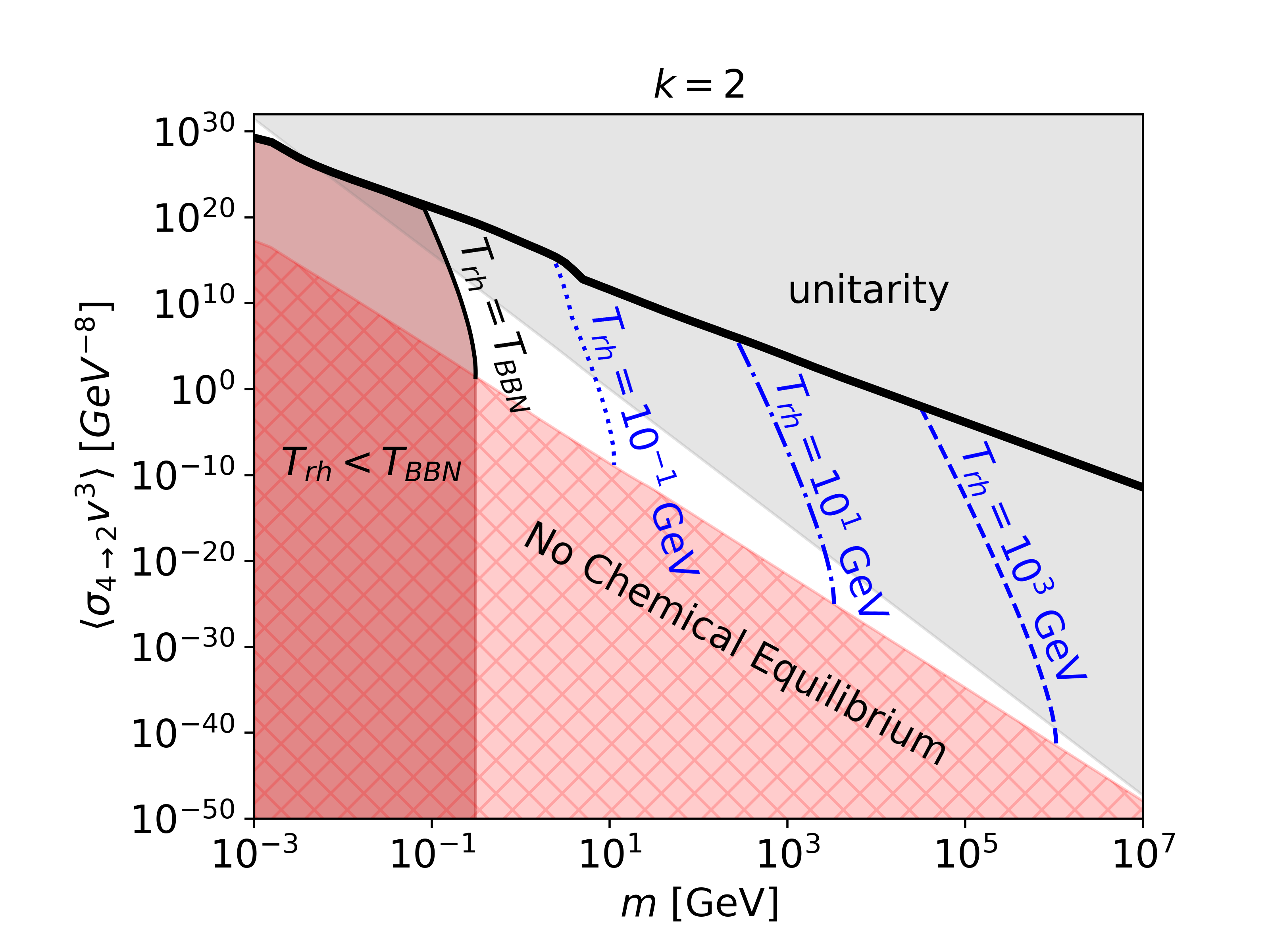}\hfill
		\includegraphics[height=4.2cm,width=5.15cm]{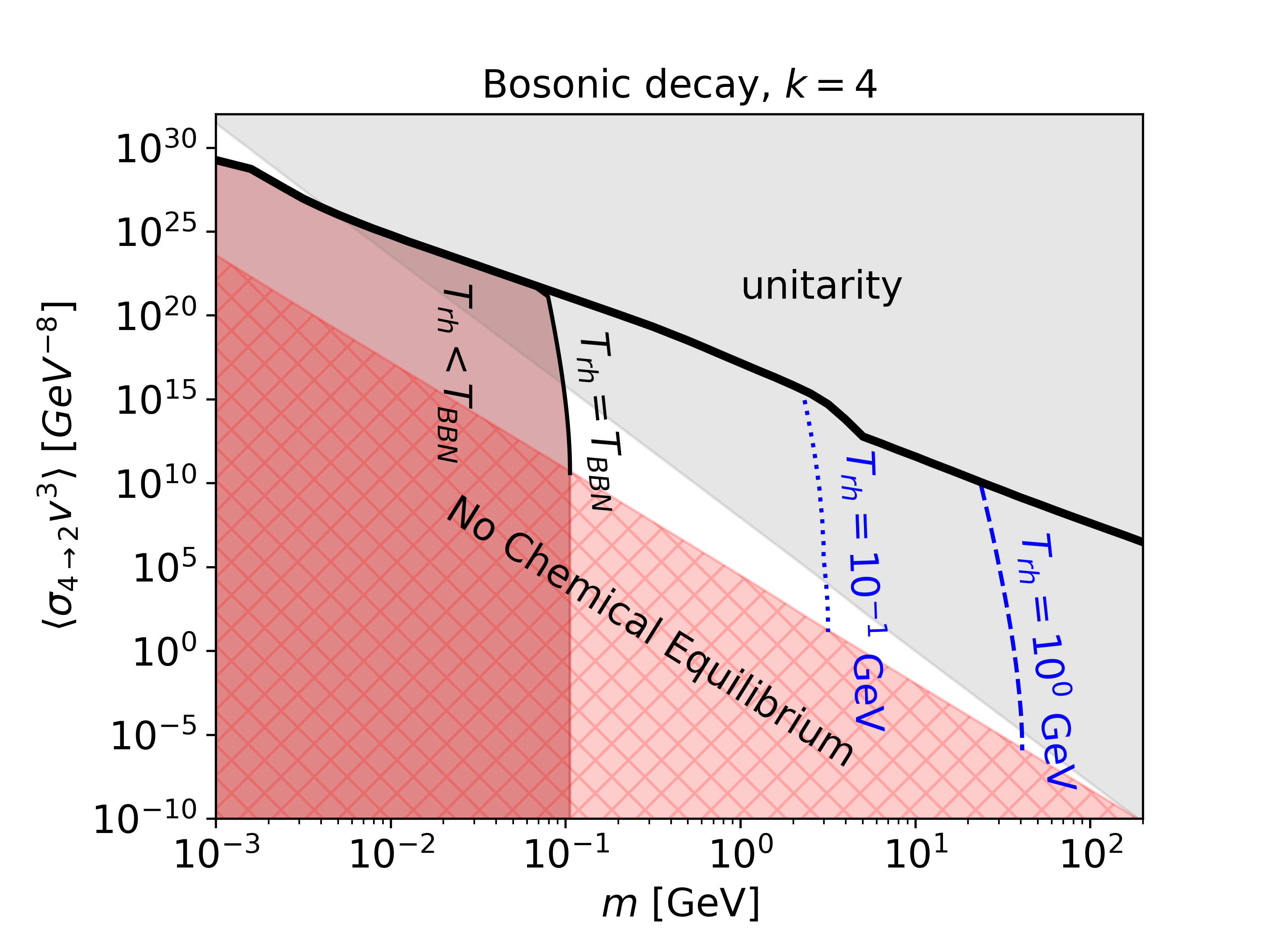}\hfill
		\includegraphics[height=4.2cm,width=5.15cm]{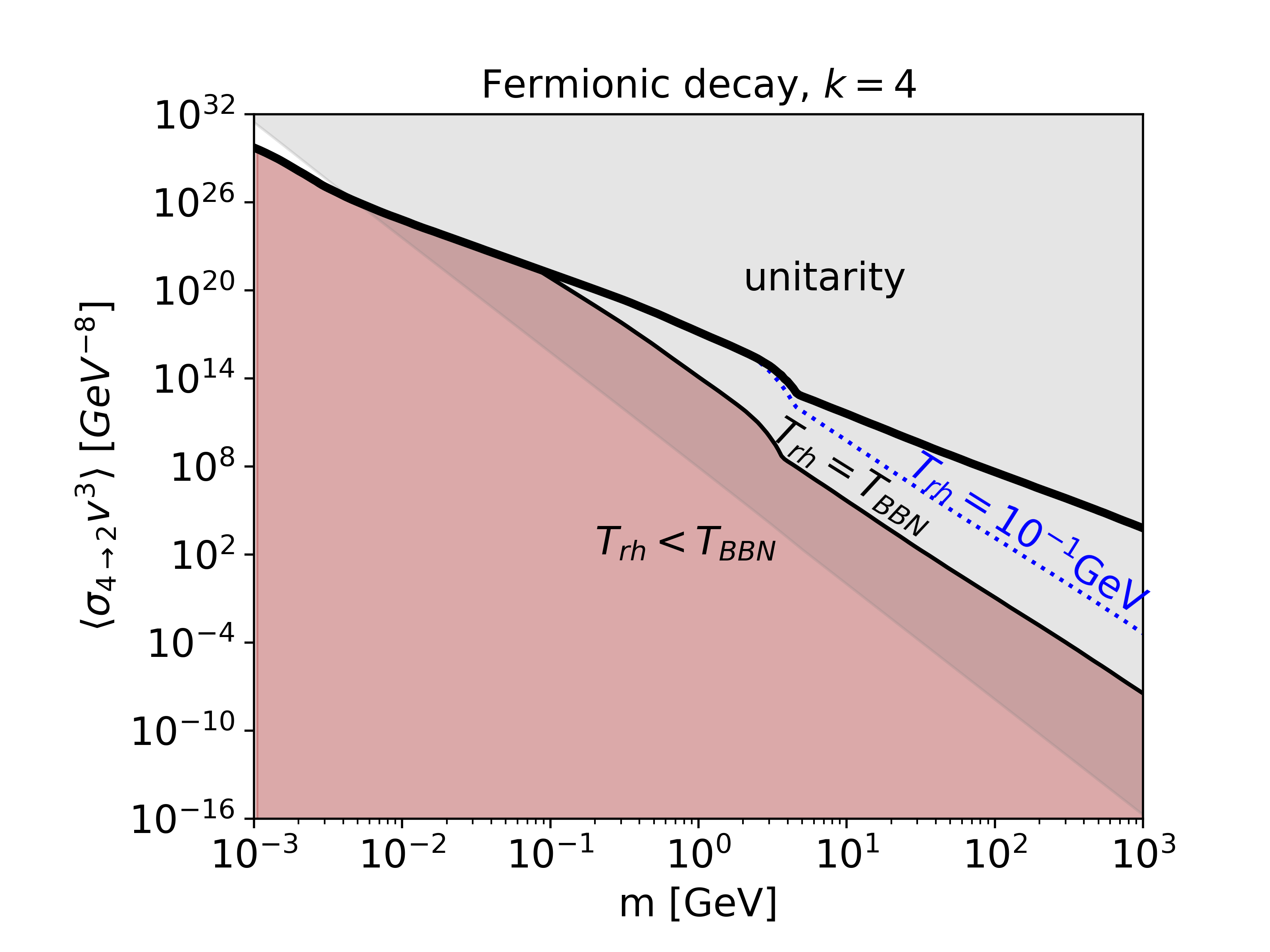}
		\caption{Same as Figure~\ref{fig:cross_32}, but for $4\rightarrow2$ process.}
		\label{fig:cross_42}
	\end{figure}
	
	\begin{figure}[H]
		\centering
		\includegraphics[height=4.2cm,width=5.15cm]{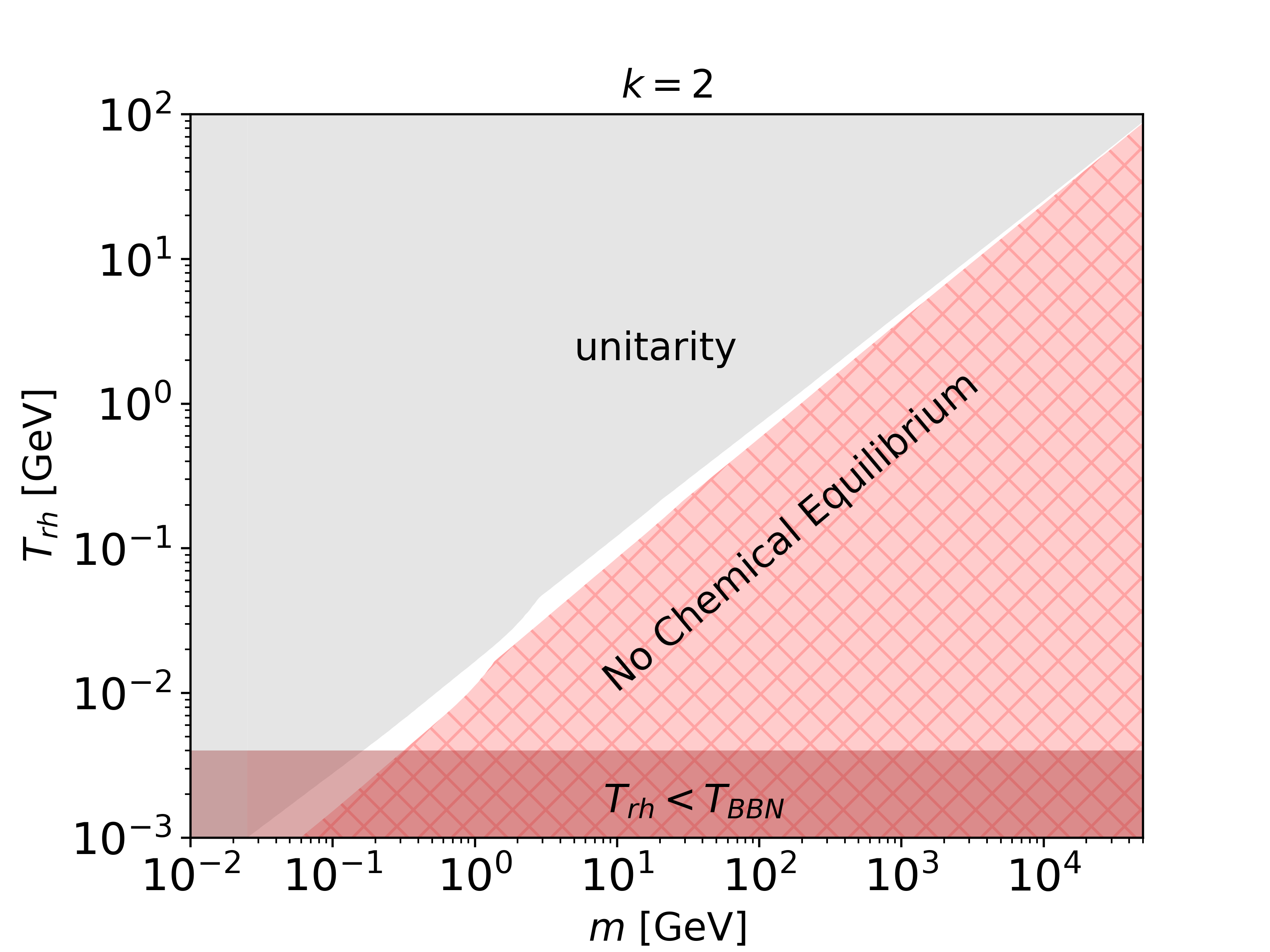}
		\includegraphics[height=4.2cm,width=5.15cm]{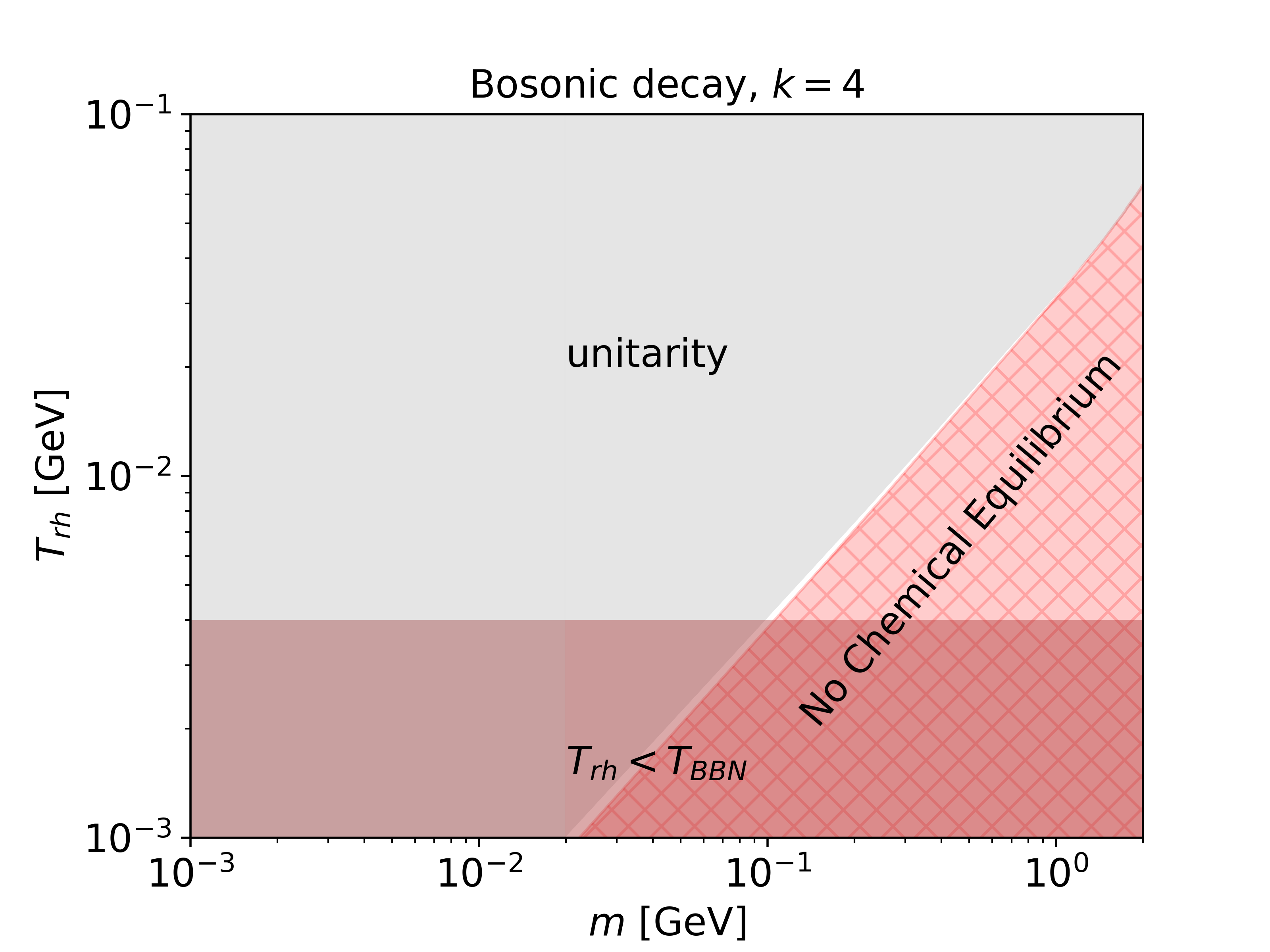}
		\caption{Same as Figure~\ref{fig:TR_m32}, but for $4\rightarrow2$ process.}
		\label{fig:TR_m42}
	\end{figure}
	 Now, the Figures~\ref{fig:xf_42}, \ref{fig:cross_42} and, \ref{fig:TR_m42} depict the equivalent results to the ones shown in Figures~\ref{fig:xf_32}, \ref{fig:cross_32} and, \ref{fig:TR_m32}, but for $4\rightarrow2$ process.
	It is important to note that the fermionic reheating picture fails to revive the parameter space, which is evident from the right plot of the lower panel of Figure~\ref{fig:cross_42}, unlike the bosonic case since temperature scales very differently with the scale factor. It is worth mentioning that the approximate mass range $\sim(7-100)$ MeV (desert region for $4\rightarrow2$ SIMP) is always forbidden by BBN constraint irrespective of the cosmological backgrounds. In passing, the fermionic reheating scenario with $k\ge4$ fails to provide the viable parameter space for $4\rightarrow2$ SIMP.	
}

\section{{Estimate of reheating temperature}}\label{realization_Trh}
{In our study, we studied the impact of fermionic and bosonic reheating on the evolution of dark matter. In the case of fermionic reheating, the inflaton decays into pair of fermions ($\psi$) through the interaction $y\phi\psi\bar\psi$ with $y$ being the Yukawa coupling where the decay rate is given by~\cite{Garcia:2020wiy}
\begin{equation}\label{decay_fer}
\Gamma_\phi^f=\frac{y_{\text{eff}}^2(k)}{8\pi}m_\phi,
\end{equation}
where
\begin{equation}\label{yeff}
y_{\text{eff}}^2(k)=y^2\alpha_y(k,\mathcal{R})\sqrt{\frac{\pi k}{2(k-1)}}\frac{\Gamma\big(\frac{k+2}{2k}\big)}{\Gamma\big(\frac{1}{k}\big)},
\end{equation}
\begin{figure}[H]
	\centering
	\includegraphics[height=6cm,width=7.2cm]{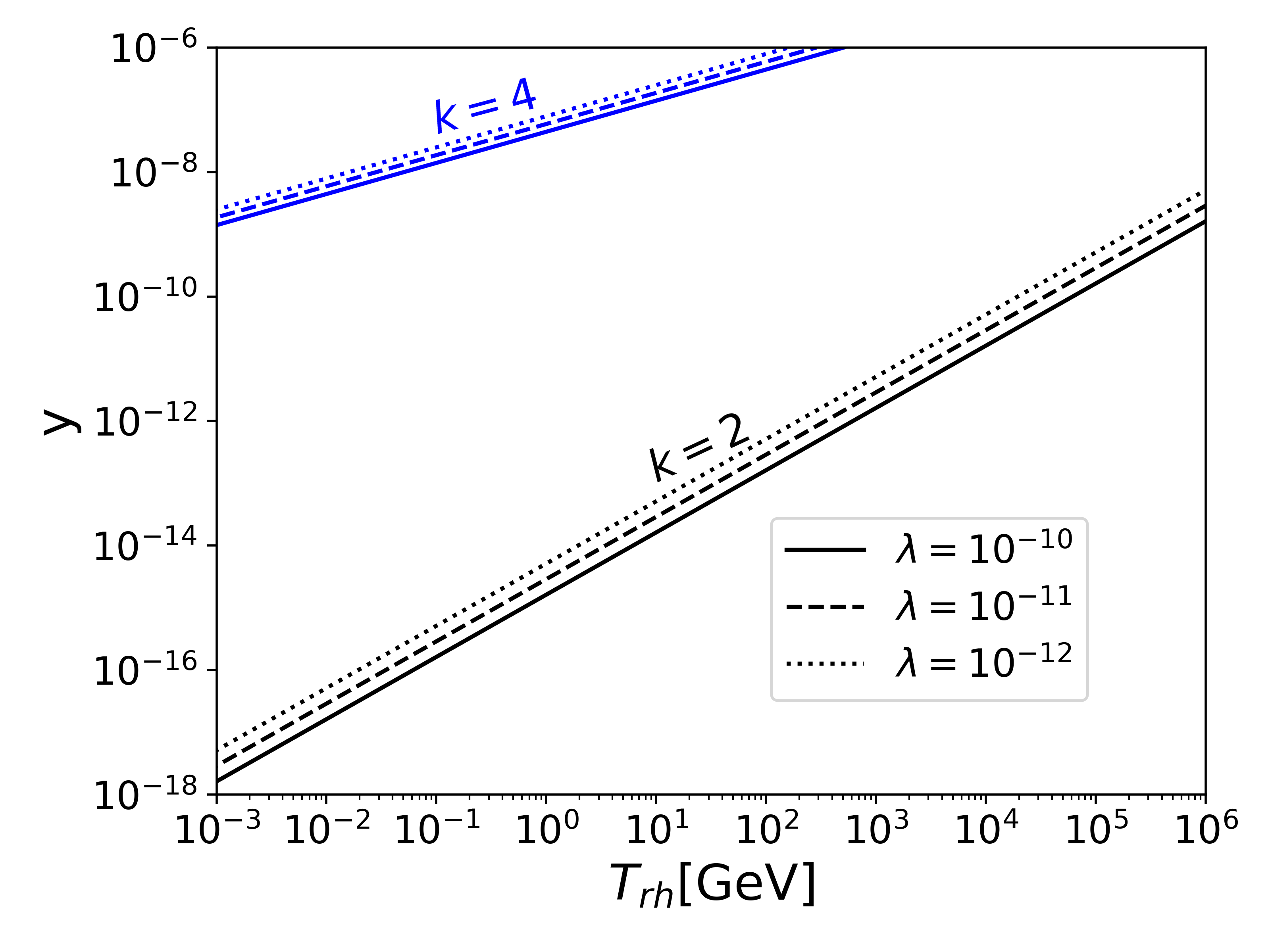}
	\includegraphics[height=5.9cm,width=7.2cm]{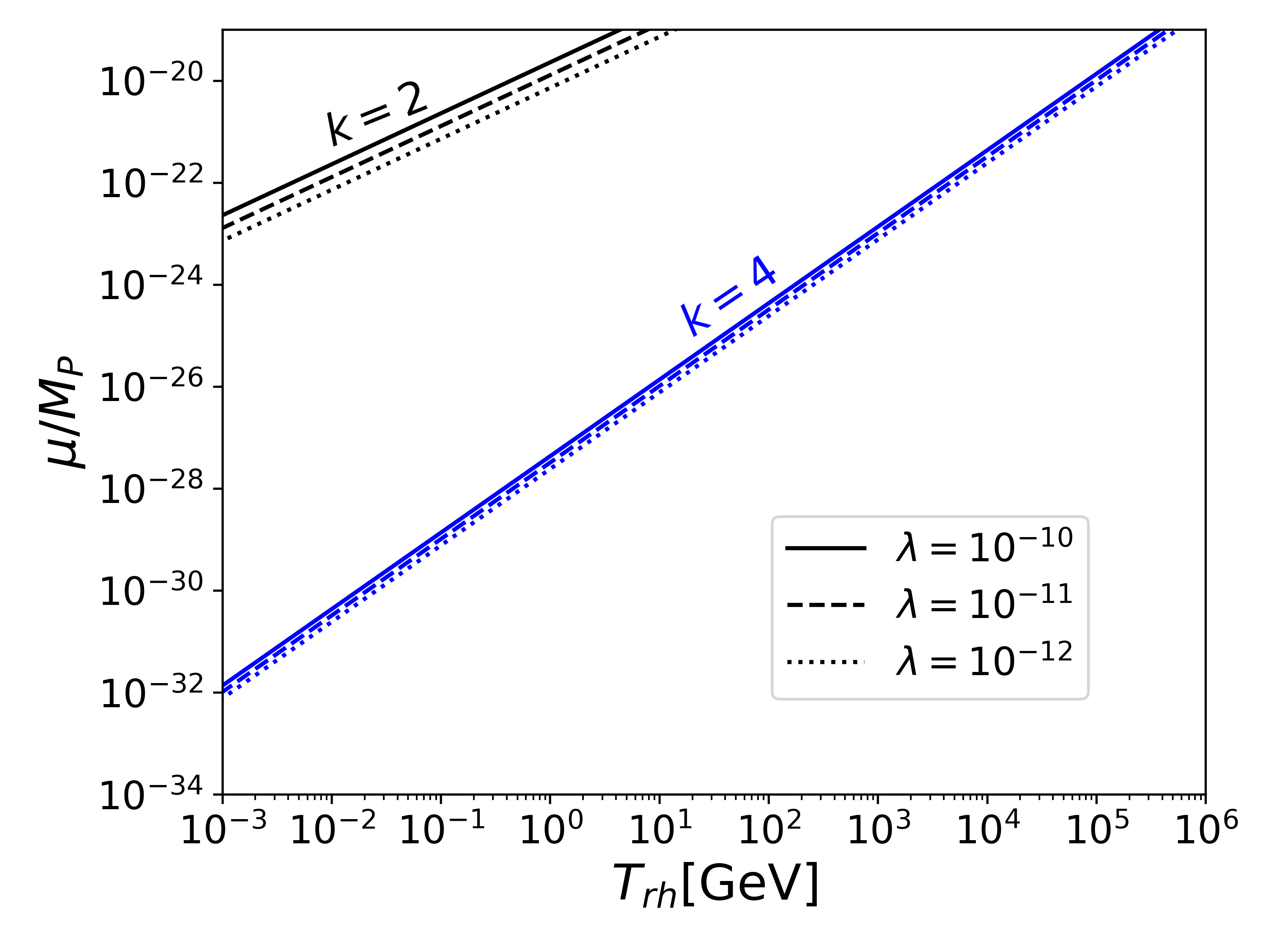}
	\caption{The left (right) plot shows the variation of Yukawa coupling ($\mu/M_P$) with the reheating temperature for three different values of the coupling $\lambda$, respectively. The black and blue lines represent the $k=2$ and $k=4$ cases, respectively.}
	\label{Tr_value}
\end{figure}
where the function $\alpha_y(k,\mathcal{R})$ depends on the kinetic factor $\mathcal{R}$ which quantifies that the inflaton induced fermion effective mass is comparable to the inflaton mass ($\mathcal{R}\geq 1$) or not ($\mathcal{R} < 1$). For $\mathcal{R}\ll1$, $\alpha_y$ changes very slowly with increasing $k$ where $\alpha_y=1$ for $k=2$ and $\alpha_y=1.05$ for $k=4$~\cite{Garcia:2020wiy} where the study is performed for $\alpha$-attractor T-model. In this work, it has shown that $\mathcal{R}<1$ for $y\leq\mathcal{O}(10^{-6})$ provided $k=4$. The reheating temperature, in this case, can be expressed as
\begin{equation}\label{Trhf}
T_{{\text{rh}},f}=\bigg(\frac{30}{\pi^2 g_*}\bigg)^{1/4}\bigg[\frac{k\sqrt{3k(k-1)}}{7-k}\lambda^{\frac{1}{k}}\frac{y_{\text{eff}}^2(k)}{8\pi}\bigg]^{k/4}M_P,
\end{equation}
In the bosonic reheating picture, the inflaton decays into two bosons (S) via a trilinear interaction $\mu\phi SS$, with $\mu$ being a dimensionful coupling where the decay width is~\cite{Garcia:2020wiy} 
\begin{equation}\label{decay_boson}
\Gamma_\phi^b=\frac{\mu_{\text{eff}}^2(k)}{8\pi m_\phi},
\end{equation}
where
\begin{equation}\label{mueff}
\mu_{\text{eff}}^2(k)=\mu^2\alpha_\mu(k,\mathcal{R})\frac{(k+2)(k-1)}{4}\sqrt{\frac{\pi k}{2(k-1)}}\frac{\Gamma\big(\frac{k+2}{2k}\big)}{\Gamma\big(\frac{1}{k}\big)},
\end{equation}
Here, the reheating temperature can be cast as~~\cite{Garcia:2020wiy}
\begin{equation}\label{Trhb}
T_{{\text{rh}},b}=\bigg(\frac{30}{\pi^2 g_*}\bigg)^{1/4}\bigg[\frac{\sqrt{3}}{8\pi(2k+1)}\lambda^{-\frac{1}{k}}\frac{\mu_{\text{eff}}^2(k)}{M_P^2}\bigg]^{k/4}M_P.
\end{equation}
Similarly, the earlier argument can be applied for the function $\alpha_\mu(k,\mathcal{R})$ where $\mathcal{R}$ quantifies that the inflaton induced boson effective mass is comparable to the inflaton mass $\mathcal{R}\geq 1$ or not $\mathcal{R} < 1$. Note that the impact of inflaton-induced effective mass is negligible for $\mathcal{R} < 1$. Here also, for $\mathcal{R}\ll1$, $\alpha_y$ is approximately constant while it only varies between $\alpha_\mu=1$ for $k=2$ and $\alpha_\mu=0.9$ for $k=4$~\cite{Garcia:2020wiy}. In ref~\cite{Garcia:2020wiy}, it has been shown that $\mathcal{R}\geq 1$ for $k=4$ whenever   $10^{-19}\leq\mu/M_P\leq 10^{-9}$, in case of $\alpha$-attractor T-model. For the reheating temperature range, ($10^{-3}$-$10^3$) GeV, $\mu/M_P$ is always less than $10^{-19}$.

The left plot of Figure~\ref{Tr_value} shows the variation of Yukawa coupling in case of fermionic reheating for $k=2$ (black lines) and $k=4$ (blue lines), respectively where the black solid, dotted and dotdashed lines represent the coupling $\lambda=10^{-10}$, $\lambda=10^{-11}$ and, $\lambda=10^{-12}$. The variation of $\mu/M_P$ with reheating temperature is showcased in the right plot of Figure~\ref{Tr_value} for the same three $\lambda$'s where the black and blue lines dictate the $k=2$ and $k=4$ scenarios, respectively. 
}
\bibliographystyle{JHEP}
\bibliography{biblio}
\end{document}